\documentclass{eas}
\usepackage{graphicx}
\def\zs{Z$_{\odot}$}
\def\ms{M$_{\odot}$}
\def\mp{M$_{\odot}$ pc$^{-2}$}

\begin{document}

\title{An Introduction to Galactic Chemical Evolution} 
\author{Nikos Prantzos}\address{Institut d'Astrophysique de Paris, UMR7095 CNRS, Université Pierre et Marie Curie }

\begin{abstract}
The formalism of the simple model of galactic chemical evolution (GCE) and its main ingredients (stellar properties, initial mass function, star formation rate and gas flows) are presented in this tutorial. It is stressed that GCE is not (yet) an astrophysical theory, but it merely provides  a framework in which the large body of data concerning the chemical composition of stars and gas in galaxies may be intrepreted. A few examples illustrating those concepts are provided, through studies of the Solar neighborhood and of the Milky Way's halo.
\end{abstract}
\maketitle
\section{Introduction}
Galactic Chemical Evolution (GCE) is the study of the evolution of the transformation  of gas into stars and of the resulting evolution of the chemical composition of a galaxy. It has not yet the status of a full astrophysical theory, like e.g. the theories of stellar evolution or of hierarchical structure formation in the universe. The reason is our poor understandig of the driver of GCE, namely large scale star formation in galaxies. On the contrary, the drivers of stellar evolution (energy producing nuclear reactions in plasmas heated by gravity) and of hierarchical structure formation (cold dark matter assembled by gravity in an expanding universe) are fairly well understood. For that reason, those theories can make relatively robust predictions, concerning for instance, various evolutionary timescales, which is not the case with GCE.

Despite that, GCE  provides a useful framework in which one may intrepret the large (and everexpanding) body of observational data concerning elemental and isotopic abundances in stars of various ages, as well as in the interstellar medium (ISM) and even the intergalactic medium (IGM).  In particular, GCE allows one to:
\begin{itemize}
\item{{\it Check/constrain our understanding of stellar nucleosynthesis} (expressed through the predicted yields of stars of various masses and metallicities) in a {\it statistical way}, i.e. by comparing GCE results to the mean trends and the dispersion of abundances and abundance ratios.}
\item{{\it Establish a  chronology of events } in a given system, by finding when the metallicity reached a certain value, or when some stellar source (SNIa, AGBs etc.) became an important contributor to the abundance of a given nuclide.}
\item{{\it Infer how a system was formed}, by constraining the history of the SFR or of various gas movements, e.g. infall for the local disk or outflow for the Galactic halo (see Sec. 3).}
\end{itemize}

Different kinds of, progressively more sophisticated, models of GCE can be developed. In the {\it simple model}, ejecta of dying stars are instantaneously mixed in the ISM, which thus aquires a unique chemical composition at any time. Models of {\it inhomogeneous GCE}  relax this ``instantaneous mixing approximation'' in a semi-analytical manner and can account for dispersion of chemical abundances in stars and the ISM. Finally {\it chemo-dynamical models} treat also the dynamics of a galaxy's gas, stars and dark matter (either in a static or in a cosmological environment)  providing a self-consistent framework for GCE (see e.g. Gibson et al. 2003 for a review). One should keep in mind, however, that (i) both the number of free parameters and the number of potentially constraining observables increase with the degree of sophistication of the model, and (ii) all models, simple and sophisticated ones,  suffer equally from our poor understanding of star formation (and stellar feedback in the case of chemo-dynamical models).

In Sec. 2 we develop the formalism of the simple GCE model (Sec. 2.1) and we present the required ingredients: stellar properties (Sec. 2.2) and yields (2.3), stellar initial mass function (2.4), star formation rate (2.5) and gaseous flows (2.6); we also provide some useful analytical solutions of the system of GCE equations, obtained under the so called "Instantaneous recycling aprproximation" (Sec. 2.7).  In  Sec. 3 we illustrate the simple model by applying it to two well observed systems: the local disk (3.1) and the Milky Way's halo (3.2).

\section{Models of GCE: formalism and ingredients}

In the simple model of GCE, a galaxy consists initially of gas of primordial composition (see contribution by A. Coc in this volume), namely by
$X_H\sim$0.75 for H and $X_{He}\sim$0.25 for $^4$He, as well as trace amounts of D, $^3$He and $^7$Li (abundances are given as {\it mass fractions} $X_i$ for element or isotope $i$). The gas is progressively turned into stars with a {\it Star Formation Rate (SFR)} $\Psi(t)$, with the star masses $M$ having a distribution $\Phi(M)$, called the {\it Initial Mass Function (IMF)}\footnote{In principle, 
the IMF may depend on time, either explicitly or implicitly (i.e. through a dependence on metallicity, which increases with time); in that case one should adopt a Star Creation Function $C(t,M)$ (making the solution of the GCE equations more difficult). In practice, however, observations indicate that the IMF does not vary with the environment, allowing to separate the variables $t$ and $M$ and adopt $C(t,M) = \Psi(t) \Phi(M)$.}. Depending on its lifetime $\tau_M$, the star of mass $M$ created at time $t$ dies at time $t+\tau_M$ and returns a part of its mass to the interstellar medium (ISM), either through stellar winds (in the case of low mass and intermediate mass stars) or through supernova explosions (in the case of massive stars\footnote{Massive stars also eject part of their mass through a wind, either in the red giant stage (a rather negligible fraction) or in the Wolf-Rayet stage (an important fraction of their mass, in the case of the most massive stars).}). The ejected material is enriched in elements synthesized by nuclear reactions in the stellar interiors, while some fragile isotopes (like D) are absent from its composition. Thus, the ISM is progressively enriched in elements heavier than H, while its D content is reduced. New stellar generations are formed from this ISM, their composition being progressively more enriched in heavy elements, i.e. with an everincreasing {\it metallicity Z} (where $Z=\Sigma X_i$ for all elements $i$ heavier than He).

In the framework of the simple model of GCE it is assumed that the {\it stellar ejecta} are immediately and efficiently mixed in the ISM\footnote{This is the so called {\it Instantaneous Mixing Approximation}, not to be confused with the {\it Instantaneous Recycling Approximation (IRA)}, to be discussed in Sec. 2.7}. As a result, the ISM is caracterized at every moment by a unique chemical composition $X_i(t)$, which is also the composition of the stars formed at time $t$. Since the surface composition of stars on the Main Sequence is not affected, in general, by nuclear reactions\footnote{An exception to that rule is fragile D, already burned in the Pre-Main Sequence all over the star's mass; Li isotopes are also destroyed, and survive only in the thin convective envelopes of the hottest stars.}, observations of stellar abundances reveal, in principle, the composition of the gas of the system at the time when those stars were formed. One may thus recover the chemical history of the system and confront observations to models of GCE.

\subsection{Formalism of the Simple Model of GCE}

The GCE scenario scetched in the previous paragraphs can be quantitavely described by a set of integrodifferential equations (see Tinsley 1980): 

The {\it evolution of the total mass of the system m(t)} is given by:
\begin{equation}
{{dm}\over{dt}} \ = \ [ \ f \ - \ o \ ]
\end{equation}
If the system evolves without any input or loss of mass, the right hand member of Eq. (1) is equal to zero; this is the so called {\it Closed Box Model}, the simplest model of GCE. The terms of the second member within brackets are optional and describe {\it infall} of extragalactic material at a rate $f(t)$ or {\it outflow} of mass from the system at a rate $o(t)$; both terms will be discussed in Sec. 2.6.

The {\it evolution of the mass of the gas} $m_G(t)$ of the system is given by:
\begin{equation}
{{dm_G}\over{dt}} \ = \ - \Psi \ + \ E \ + \ [ \ f \ - \ o \ ]
\end{equation}
where $\Psi(t) $ is the Star Formation Rate (SFR)  and $E(t)$ is the {\it Rate of mass ejection by dying stars}, given by:
\begin{equation}
E(t) \ = \ \int_{M_t}^{M_U} \ (M-C_M) \ \Psi(t-\tau_M) \ \Phi(M) \ dM
\end{equation}
where the star of mass $M$, created at the time $t-\tau_M$, dies at time $t$ (if $\tau_M<t$) and leaves a Compact object (white dwarf, neutron star, black hole) of mass $C_M$, i.e. it ejects a mass $M-C_M$ in the ISM. The integral in Eq. (3) is weighted by the Initial Mass Function of the stars $\Phi(M)$ and runs over all stars heavy enough to die at time $t$, i.e. the less massive of them has a mass $M_t$ and a lifetime  $\tau_M\leq t$. The upper mass limit of the integral $M_U$ is the upper mass limit of the IMF and is discussed in Sec. 2.4.

Obviously, the {\it mass of stars} $m_S(t)$ of the system can be derived through:
\begin{equation}
m \ = \ m_S \ + \ m_G
\end{equation} 
As time goes on, a progressively larger part of the mass of stars is found in the form of {\it compact objects  of total mass} $m_C(t)$, while the {\it mass of stars still shining (live stars)} is $m_L(t)$; one has then:
\begin{equation}
m_S \ = \ m_L \ + \ m_C
\end{equation}
The mass of compact objects is
\begin{equation}
m_C \ = \ \int_0^t \ c(t) \ dt
\end{equation}
where $c(t)$ is the {\it mass creation rate of compact objects}, given by:
\begin{equation}
c(t) \ = \ \int_{M_t}^{M_U} \ C_M \ \Psi(t-\tau_M) \ \Phi(M) \ dM
\end{equation}
The mass of live stars $m_L(t)$ can then be calculated by using Eqs. 2.5 to 2.7.

The evolution of the chemical composition of the system is decribed by equations similar to Eqs. (2.2) and (2.3). The {\it mass of element/isotope i} in the gas is $m_i=m_GX_i$ and its evolution  is given by:
\begin{equation}
{{d(m_G \ X_i)}\over{dt}} \ = \ - \Psi X_i \ + \ E_i \ + \ [ \ f X_{i,f} \ - \ o X_{i,o} \ ]
\end{equation}
i.e. star formation at a rate $\Psi$ removes element $i$ from the ISM at a rate $\Psi X_i$, while at the same time stars re-inject in the ISM that element
at a rate $E_i(t)$. If infall is assumed, the same element $i$ is added to the system at a rate $f X_{i,f}$, where $X_{i,f}$ is the abundance of nuclide $i$  in the infalling gas (usually, but not necessarily, assumed to be primordial). If outflow takes place, element $i$ is removed from the system at a rate $o X_{i,o}$ where $X_{i,o}$ is the abundance in the outflowing gas; usually, $X_{i,o}$=$X_i$, i.e. the outflowing gas has the composition of the average ISM, but in some cases it may be assumed that the hot supernova ejecta (rich in metals) leave preferentially the system, in which case $X_{i,o} > X_i$ for metals.

The {\it rate of ejection of element i by stars} is given by:
\begin{equation}
E_i(t) \ = \ \int_{M_t}^{M_U} \ Y_i(M) \ \Psi(t-\tau_M) \ \Phi(M) \ dM
\end{equation}
where $Y_i(M)$ is the {\it stellar yield of element/isotope i}, i.e. the mass ejected in the form of that element by the star of mass $M$. Note that $Y_i(M)$ may depend {\it implicitly} on time $t$, if it is metallicity dependent (see Sec. 2.3.2).

The masses involved in the system of Eqs. 2.1 to 2.9 may be either {\it physical } masses, i.e. $m$, $m_G$, $m_S$ etc. are expressed in \ms \ and $\Psi(t)$, $E(t)$, $c(t)$ etc. in \ms \ Gyr$^{-1}$, or {\it reduced} masses ({\it mass per unit final mass of the system}), in  which case $m$, $m_G$, $m_S$ etc. have no dimensions and  $\Psi(t)$, $E(t)$, $c(t)$ etc. are in  Gyr$^{-1}$. The latter possibility allows to perform calculations for a system of arbitrary mass and normalise the results to the known/assumed mass of that system; note that instead of mass, one may use volume or surface mass densities.

The system of integrodifferential equations 2.1 to 2.9 can only be solved numerically, unless some specific assumptions are made, in particular the  Instantaneous recycling aproximation (IRA, to be discussed in Sec. 2.7). Its solution requires three types of ingredients:

\begin{itemize}
\item  {\it Stellar properties}: stellar lifetimes $\tau_M$, masses of stellar residues $C_M$
and yields $Y_i(M)$; all those quantities can be derived from the theory of stellar evolution and nucleosynthesis and depend (to various degrees) on the initial stellar metallicity Z. 
\item  {\it Collective stellar properties}: the initial mass function $\Phi(M)$ and the Star Formation Rate $\Psi$; none of them can be reliably derived from first principles at present, and one has to rely on empirical prescriptions.
\item  {\it Gas flows} into and out of the system (infall, inflow, outfow, wind): in simple GCE models these factors are optional, i.e. their introduction depends on the nature of the considered galactic system  (e.g. infall for the solar neighborhood or winds for small galaxies). In more physical  (e.g. hydrodynamical) models, they should stem naturally from the physics of the system.
\end{itemize} 
We discuss those ingredients in the following  sections.

\begin{figure}
\begin{center}
 \includegraphics[width=5.8cm,angle=-90]{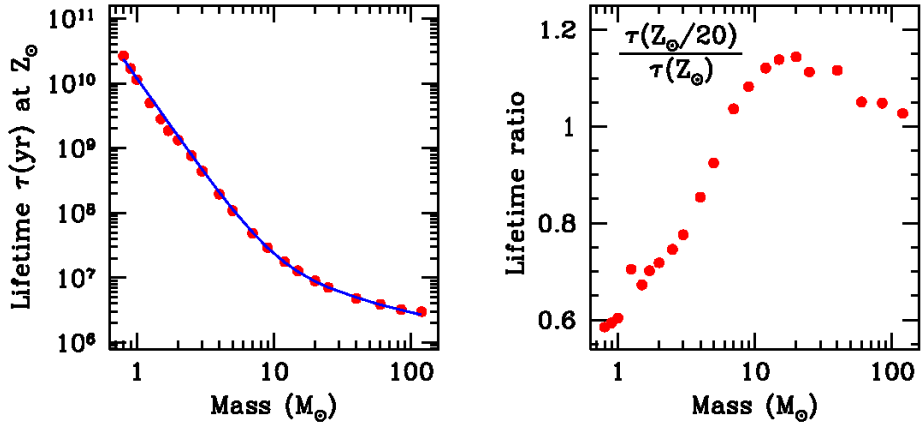}
\caption{{\it Left}: Lifetimes of stars of solar metallicity \zs; {\it points} are from Geneva stellar models (Schaller et al. 1992) and the {\it curve} is a fit to those points (see Equ. 2.10). {\it Right}: Ratio of stellar lifetimes at \zs/20 and \zs \ for the same models.}
\end{center}
\end{figure}

\subsection {Stellar lifetimes and residues}

Stellar lifetime is a strongly decreasing function of stellar mass (see Fig. 1). Its precise value depends on the various assumptions (about e.g. mixing, mass loss, etc.) adopted in stellar evolution models and, most importantly on stellar metallicity. Indeed, low metallicity stars have lower opacities and are more compact and hot than their high metallicity counterparts; as a result, their lifetimes are shorter (see Fig. 1 right). However, in stars with M$>$2 \ms, where  H burns through the CNO cycles, this is compensated to some degree by the fact that the H-burning rate (proportional to the CNO content) is smaller, making the corresponding lifetime longer; thus, for M$>$10 \ms, low metallicity stars live slightly longer than solar metallicity stars. Of course, these results depend strongly on other ingredients, like e.g. rotation (see Meynet, this volume). In principle, such variations in $\tau_M$ should be taken into account in GCE models; in practice, however, the errors introduced by ignoring them are smaller than the other uncertainties of the problem, related e.g. to stellar yields or to the IMF\footnote{Metallicity dependent lifetimes {\it have to be taken into account} in models of the spectro-photometric evolution of galaxies, where they have a bigger impact.}. 

The lifetime of a star of mass M (in \ms) with metallicity \zs \ can be aproximated by:
\begin{equation}
\tau(M) \ = \ 1.13 \ 10^{10} M^{-3} \ + \ 0.6 \ 10^8 M^{-0.75} \ + \ 1.2 \ 10^6  \ {\rm yr}
\end{equation}
This fitting formula is displayed as solid curve in Fig. 1 (left). A \zs \ star of 1 \ms, like the Sun, is bound to live for 11.4 Gyr, while a 0.8 \ms \ star for $\sim$23 Gyr; the latter, however, if born with a metallicity Z$\leq$0.05 \zs, will live for ``only'' 13.8 Gyr, i.e. its lifetime is comparable to the age of the Universe. Stars of mass 0.8 \ms \ are thus the lowest mass stars that have ever died since the dawn of time (and the heaviest stars surviving in the oldest globular clusters).

The masses of stellar residues are derived from stellar evolution calculations, confronted to observational constraints. In the regime of Low and Intermediate Mass Stars (LIMS\footnote{LIMS are defined as those stars evolving to white dwarfs. However, there is no universal definition  for the mass limits characterizing Low and Intermediate Mass stars. The upper limit is usualy taken around 8-9 \ms, although values as low as 6 \ms \ have been suggested (in models with very large convective cores). The limit between Low and Intermediate masses is the one separating stars powered on the Main Sequence by the p-p chains from those powered by the CNO cycle and is $\sim$1.2-1.7 \ms, depending on metallicity.}), i.e. for M/\ms$\leq$8-9, the evolutionary outcome is a white dwarf (WD), the mass of which (in \ms) is given by (Weidemann 2000):
\begin{equation}
C_M (WD) \ = \ 0.08 \ M \ + \ 0.47  \ \ \ \ \ \  \ \ \ \ \ \ \ \ \ \ \  \ \ \ \ \ (M<9)
\end{equation}
Thus, the heaviest WD has a mass of $\sim$1.2 \ms, slightly below the Chandrasekhar mass limit of $\sim$1.44 \ms. The less massive WDs formed from single mass stars have a mass $\sim$0.53 \ms. Observations show indeed that WD masses cluster around that value. Note that WD with much lower masses have been found (even below 0.2 \ms, e.g. Kilic et al. 2006), but they result from binary system evolution.

\begin{figure}
\begin{center}
 \includegraphics[width=5.8cm,angle=-90]{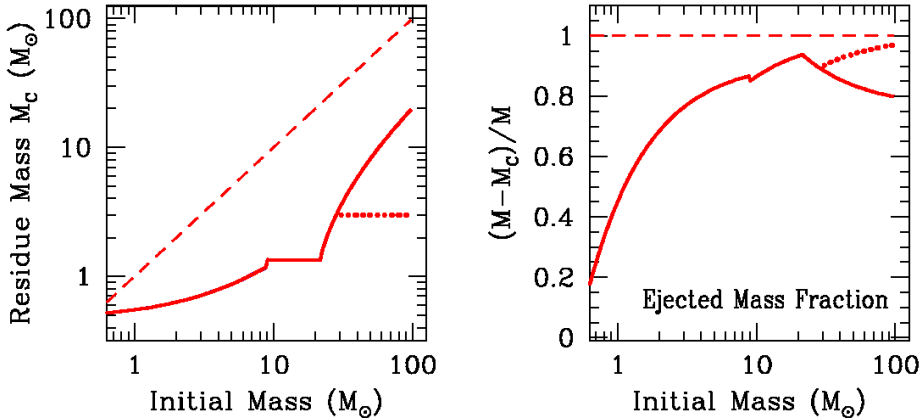}
\caption{{\it Left:} Masses of stellar residues as a function of initial stellar mass, for stars of metallicity \zs; for massive stars (M$>$30 \ms)  the two curves correspond to different assumptions about mass loss, adopted in Limongi and Chieffi ({\it solid curve}, this volume) and Woosley and Heger (2007, {\it dotted curve}), respectively. {\it Right:}   Mass fraction of the ejecta as a function of initial stellar mass; the two curves for M$>$30 \ms \ result from the references in the left figure. }
\end{center}
\end{figure}

Stars more massive than  8-9 \ms \ explode as supernovae (SN), either after electron captures in their O-Ne-Mg core (M$\leq$11 \ms) or after Fe core collapse (M$\geq$11 \ms). The nature and mass of the residue depends on the initial mass of the star and on the mass left to it prior to the explosion. It is often claimed that solar metallicity stars of M$\leq$25 \ms \ leave behind a Neutron Star (NS), while heavier stars leave a black hole (BH). Neutron star masses are well constrained by the observed masses of pulsars in binary systems: $M_{NS}$=1.35$\pm$0.04 \ms \ (Thorsett and Chakrabarty 1999), which is adopted here
\begin{equation}
C_M(NS) \ = \ 1.35 \ \ \ \ \ \ \ \ \ \ \ \ \ \ \ \ \ \ \ \ \  \ \ \ \ \ (9 < M < 25)
\end{equation}
i.e. $C_M$ is independent of the initial mass $M$ in that case.  However, black hole masses are not known observationally as a function of the progenitor mass, while theoretical models are quite uncertain in that respect. Thus, for solar metallicity stars with mass loss, the Frascati models (Limongi, this volume) produce black hole masses proportional to the initial mass, which can be approximated (in \ms) by:
\begin{equation}
C_M(BH) \ = \ 0.24 \ M  \ - 4 \ \ \ \ \ \ \ \ \ \ \ \ \ \ \ \ \ \ \ \ \ \  \ \ \ \ \ (M > 25)
\end{equation}
while Woosley and Heger (2007) find that, as a result of high mass losses for M$\geq$25 \ms, the average black hole mass (in \ms) is
\begin{equation}
C_M(BH) \ = 3  \ \ \ \ \ \ \ \ \ \ \ \ \ \ \ \ \ \ \ \ \ \ \ \ \ \ \ \ \ \ \ \ \ \ (M > 25)
\end{equation}
(see Fig. 2).  It is commonly accepted that  black hole masses are much larger at low metallicities, where the effects of mass loss are negligible. Note, however, that the magnitude of the effect could be moderated in models with rotational mixing, which induces mass loss even at very low metallicities (see Meynet, this volume). 

It should be stressed that the mass limit between the progenitors of neutron stars 
and black holes is not understood at present, and adopting simple theoretical ``recipes'' can be strongly misleading. For instance, Muno (2006) points out that a magnetar (strongly magnetised neutron star) is found in a star cluster less than 4.5 Myr old (as deduced from the absence of red supergiants in it), implying a progenitor mass larger than 50 \ms. Of course, it is possible that strong mass loss in a binary system reduced the mass of such a heavy progenitor to a very low level prior to the explosion.

\subsection{Stellar yields}

\subsubsection{Definitions}

The quantities required in Eq. 2.9 are the {\it stellar yields} $Y_i(M)$, representing the mass ejected in the form of element $i$ by a star of mass $M$. Those quantities are obviously 
$Y_i(M)\geq$0 ($Y_i$=0 in the case of an isotope totally destroyed in stellar interiors, e.g. deuterium). However, their value is of little help in judging whether star $M$ is an important producer of isotope $i$ (e.g. by knowing that a 20 \ms \ star produces 10$^{-3}$ \ms \ of Mg or 1 \ms \ of O, one cannot judge whether such a star contributes significantly - if at all - to the galactic enrichmemnt in those elements). 

More insight in that respect is obtained through the {\it net yields} $y_i(M)$, which represent the {|it newly created mass of nuclide } $i$ from a star, i.e.
\begin{equation}
y_i(M) \ = \ Y_i(M) - \ M_{0,i}(M)
\end{equation}
where $ M_{0,i}(M)$ is the mass of nuclide $i$ originally present in the part of the star that is finally ejected:
\begin{equation}
 M_{0,i}(M) \ = \ X_{0,i} (M \ - \ C_M)
\end{equation}
and $X_{0,i}$ is the mass fraction of nuclide $i$ in the gas from which the star is formed. Obviously, $y_i(M)$ may be positive, zero or negative, depending on whether star $M$ creates, simply re-ejects or destroys nuclide $i$. Net yields {\it are not used} in numerical models of GCE, but {\it they are used} in analytical models, adopting the Instantaneous recycling approximation (see Sec. 2.7).

Finally, the {\it production factors} $f_i(M)$ are defined as:
\begin{equation}
f_i(M) \ = \  {{Y_i(M)}\over{M_{0,i}(M)}}
\end{equation}
They are useful in the sense that they immediately indicate whether star $M$ is an important producer of nuclide $i$. For instance, massive stars are the exclusive producers of oxygen, for which  $f\sim$10 on average (see Fig. 4). If such stars produce another  nuclide $L$ with, say, $f\sim$3 only, they are certainly important contributors, but they cannot account for the solar $L/O$ ratio; another source is then required\footnote{The example is taken on the case of iron, for which another source is required, beyond massive stars; that source is SNIa (see Sec. 2.3.5)} for nuclide $L$.
Note that the use of production factors as defined in Eq. 2.17 and described in the previous paragraph is interesting only when comparison is made for a star of a given initially metallicity (see e.g. Fig. 4, where denominators in Eq. 2.17 are calculated for \zs \ always, even if yields are for different initial metallicities).

   The properties of the various quantities defined in this section are summarized in Table 1.

\begin{table}
\caption{Yield definitions}
\label{tabl1}
\begin{tabular}{rccc}
\hline
\hline
Nuclide $i$ & Yields $Y_i(M)$ & Net yields $y_i(M)$  & Production factors $f_i(M)$  \\
\hline
Created     & $>M_{0,i}$ &  $>$ 0  &  $>$ 1 \\
Re-ejected  & = $M_{0,i}$ &  = 0  &  = 1 \\
Destroyed   & $< M_{0,i}$ &  $<$ 0  &  $<$ 1 \\
\hline
\hline
\end{tabular}
$M_{0,i}$ is defined in Eq. 2.16.
\end{table}

\subsubsection{Metallicity effects: primary vs. secondary and odd vs. even nuclides}

Nuclides with yields independent of the initial metallicity of the star are called {\it primary}.
They are produced exclusively from the initial H and He of the star 
(the abundances of which vary little throughout  galactic history). This is, in principle, the case for most isotopes up to the Fe peak, as well as the r-isotopes.
In stellar models with no mass loss, such as those constructed until recently (including the widely used Woosley and Weaver 1995 models) initial metallicity plays a negligible role in the overall structure of the star, and thus the yields of primary elements (like e.g. oxygen or calcium) correspond to their definition, i.e. they are metallicity independent. However, in models with mass loss the situation is different. Nuclides produced early in the lifetime of the star (i.e. during H and He burning), like C and O, are directly affected by mass loss: for instance, large amounts of material are ejected in the form of He and C in the Wolf-Rayet phase of high metallicity stars, leaving a small core available to turn into O; on the contrary, lower metallicity stars produce more O and less He and C (see Meynet, this volume). Thus, the yields of those nuclides, although defined as primaries, show a considerable dependence on metallicity. Nuclides produced in later evolutionary phases, much closer to the stellar center (like Si and Ca) are less affected by mass loss. Note that rotational mixing, inducing mass loss even at low metallicities, makes the mass loss effect less metallicity dependent.

Nuclides with yields dependent on the abundance of some non-primordial nucleus in the original stellar composition are called {\it secondary}. This is (again, in principle !) the case of : 
s-nuclides, produced by neutron captures on initial Fe-peak nuclei ;  $^{14}$N and $^{13}$C in moderately massive stars (made from initial C and O in the CNO cycle of H-burning); $^{18}$O, made by $^{14}$N($\alpha$,$\gamma$)$^{18}$F($\beta^+$)$^{18}$O in early He-burning inside massive stars. Note that in the latter case $^{18}$O is made {\it not} from the initial $^{14}$N of the star (itself a secondary nuclide), because in that case $^{18}$O would be a {\it tertiary} nuclide; it is made from the $^{14}$N already synthesised in the star through the CNO cycle, i.e. it is essentially made from initial C and O. In fact, we know of no {\it tertiary} nuclide, made from a secondary one initially present in the star. For instance, p-nuclides are made from the photo-desintegration of s-nuclei (see Arnould, this volume) which are produced inside the star from the initial Fe (a primary nuclide); thus, p-nuclides are  also secondary.

Most of the metals in a massive star (namely C and O) turn into $^{14}$N in H-burning and
most of that $^{14}$N  turns into $^{22}$Ne during He-burning, through the reaction chain leading to $^{18}$O (see previous paragraph) plus $^{18}$O($\alpha$,$\gamma$)$^{22}$Ne. The beta-decay involved in this chain modifies the neutron-to-proton ratio of the star, since the initial mixture of $^{12}$C and $^{16}$O has  the same number of neutrons N and protons Z (its {\it neutron excess} $\eta=\frac{N-Z}{N+Z}$ is $\eta$=0) while $^{22}$Ne (Z=10 and N=12)  has $\eta$=0.09. This, metallicity dependent, surplus of neutrons affects the products of subsequent burning stages: odd nuclides like e.g. $^{23}$Na, $^{25}$Mg, $^{27}$Al etc. are more readily produced in large $\eta$ (large initial metallicity) environments with respect to even nuclides; this the so-called {\it odd-even effect} (Arnett 1996). For such nuclides, use of {\it metallicity-dependent yields} is mandatory for GCE studies.


\begin{figure}
\begin{center}
 \includegraphics[width=8.5cm,angle=-90]{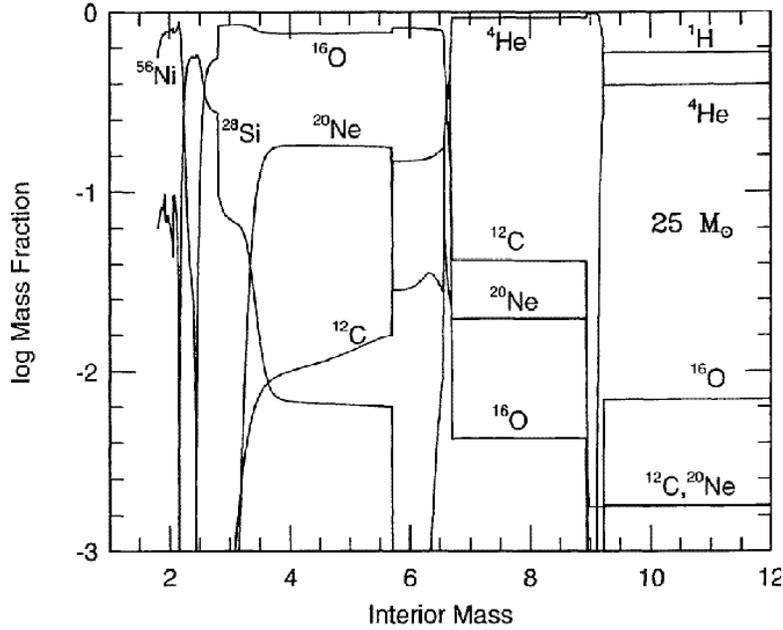}
\caption{Interior composition of a 25 \ms \ star after its explosion; only major isotopes are displayed (from Woosley and Weaver 1995).}
\end{center}
\end{figure}

\subsubsection{Yields of massive stars} 

Massive stars produce the quasi-totality of nuclides between carbon and the Fe-peak, as well as most of  the heavier than iron nuclei: the light s- (up to Y), the r- (neutron-rich) and p- (neutron poor, w.r.t. to the nuclear stability valley) nuclides.

The typical composition profile of a massive star after its supernova explosion is shown in Fig. 3. It displays the typical onion-skin structure, with heavier elements encountered in its various layers from the surface to the center. The thickness of the layers depends on assumptions about convection and other mixing processes. The abundances of the nuclides in each layer depend essentially on adopted nuclear reaction rates (but also on the mixing processes). The abundances in the outer layers depend also on the adopted mass loss rates (combined to the mixing processes). The abundances in the inner layers depend also on assumptions about the explosion of the star (mass cut, energy, fallback etc.). In particular, intermediate mass elements produced in massive stars may be 
divided in three main classes: \par
- In the 1st class belong N, C, O, Ne and Mg, which are mainly produced 
in hydrostatic burning phases and are found mainly in layers that are not
heavily processed by explosive nucleosynthesis; the yields of those elements
depend on the pre-supernova model (convection criterion, mixing processes,
mass loss and nuclear reaction rates). In particular, rotational mixing may produce primary N, even in very low metallicity stars (see Meynet, this volume).\par
- In the 2nd class belong Al, Si, S, Ar and Ca. 
They are produced by hydrostatic
burning, but their abundances are substantially affected by the passage of 
the shock wave. Their yields depend on both the pre-supernova model and the
shock wave energy. \par
- In the 3d class belong the Fe-peak nuclei, as well as some lighter elements
like Ti; their yields depend crucially upon the explosion mechanism and the
position of the ``mass-cut''.  

 In view of the important uncertainties still affecting the ingredients of stellar physics, it is obvious that stellar yields have large uncertainties.
How could the validity of the theoretical stellar yields be checked?
Ideally, individual yields should be compared to abundances measured in 
supernova remnants of stars with known initial mass and metallicity!
However, such 
opportunities are extremely  rare. In the case of SN1987A, theoretical 
predictions for a 20 \ms \ progenitor are in rather good agreement 
with observations of C, O, Si, Cl and Ar (Thielemann et al. 1996). 
SN1987A allowed also to ``calibrate'' the Fe yield 
($\sim$0.07 \ms) from the optical light curve (powered at late times from
the decay of $^{56}$Co, the progeny of $^{56}$Ni), extrapolated
to the moment of the explosion (e.g. Arnett et al. 1989). 
This may be the best way to evaluate 
the Fe yields of other SNII at present,
until a convincing way of 
determining the ``mass-cut'' from first principles emerges.

\begin{figure}
\begin{center}
 \includegraphics[width=9.cm,angle=-90]{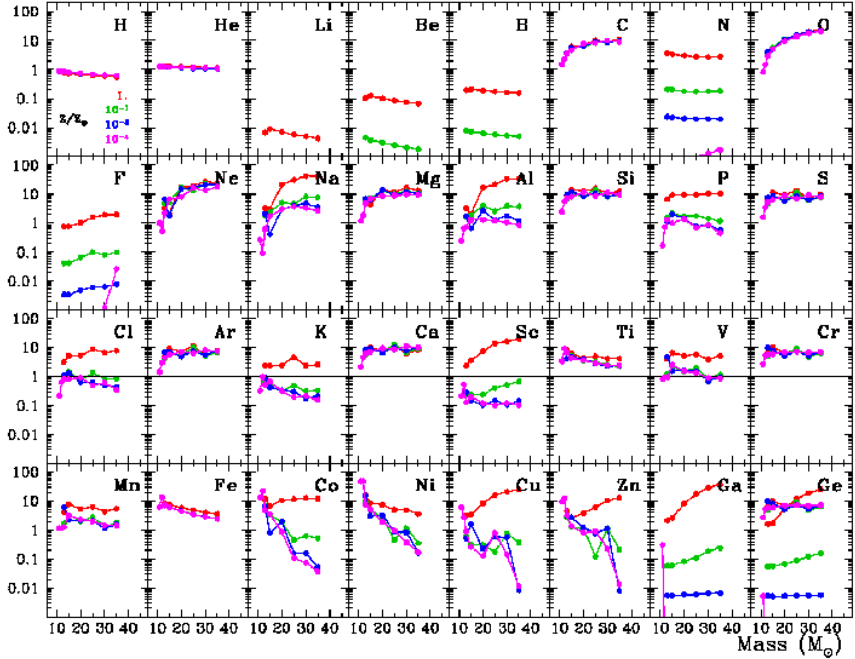}
\caption{Production factors $f_i(M) = \frac{Y_i(M)}{X_{\odot,i}(M-C_M)}$ for elements up to Ge in the ejecta of stars with no mass loss, of initial metallicity  $Z$/\zs=1, 0.1, 0.01 and 0.0001 (upper left panel) in the mass range 12-35 \ms \ (data from Chieffi and Limongi 2004). Primary elements (C, O, Ne, Mg, Si, S, Ar, Ca, Ti, Cr, Fe) have metallicity independent $f$. N is produced only as secondary in massive stars with no rotational mixing. The odd-even effect is clearly seen for Na, Al, P, Cl, K, Sc, V, Mn,while Li, Be and B are depleted to various degrees. A horizontal dotted line at $f$=1 in all panels indicates ejection of material with solar abundance.}
\end{center}
\end{figure}

Our confidence to massive star yields stems rather from the fact that, when averaged over a reasonable IMF and used in a GCE model with a SFR appropriate for the solar neighborhood, they reproduce the solar composition between C and Fe-peak within a factor of 2-3 (see Sec. 3.1.2). This is a remarkable achievement, taking into account that solar abundances in that mass range vary by  factor of $\sim$10$^6$ between the most abundant of them (O) and the less abundant one (Sc). A similar situation holds for the light s-nuclides and the p-nuclides (both products of massive stars), whereas the situation with r-nuclide yields is problematic at present (see Arnould, this volume).

\subsubsection{Yields of intermediate mass stars}

Intermediate mass stars synthesize substantial amounts of several nuclides, mainly in the Asymptotic Giant Branch (AGB) phase, when H and He burn intermittently in two shells surrounding the inert CO core. Nucleosynthesis occurs in those shells, as well as in the bottom of the convective AGB envelope, if it penetrates in regions of high enough temperature ({\it Hot Bottom Burning} or HBB). Such stars are the main producers of heavy s-nuclei at a galactic level and they synthesize large  amounts of $^4$He, $^{14}$N, $^{13}$C, $^{17}$O, $^{19}$F. etc. (see Siess, this volume). In fact, they were thought to be the main producers of those light nuclides.  However, massive stars can also produce large amounts of those nuclides  (especially when  rotational mixing is taken into account) and eject them through stellar winds (see Meynet, this volume); in view of the uncertainties affecting the yields of both AGBs and massive stars, it is not clear yet whether the contribution of AGBs  to the solar abundances of those nuclides is  predominant or not. For instance, it has been suggested that massive stars are the main source of carbon in the Milky Way (Prantzos et al. 1994).

Yield uncertainties are, in general,  more important in the case of AGBs than in the case of massive stars. The role of mixing is indeed critical, in several respects: for providing the appropriate neutron source in the case of the s-process (i.e. through $^{13}$C($\alpha$,n) in the intershell region, see e.g. Goriely and Siess 2006); for bringing material from the He-shell up to the envelope (the so-called {\it third dredge-up});   and for bringing the bottom of the convective envelope to sufficiently high temperatures as to burn hydrogen through the CNO cycle (HBB, synthesizing primary $^{14}$N and $^{13}$C). Mass loss is also critical, since it determines the duration, and the degree of nuclear processing, of the AGB phase.
Note that, due to the complexity of the AGB phase, yields are often calculated by "post-processing", i.e. nucleosynthesis is calculated in an envelope decoupled from the stellar core.

\begin{figure}
\begin{center}
 \includegraphics[width=9cm,angle=-90]{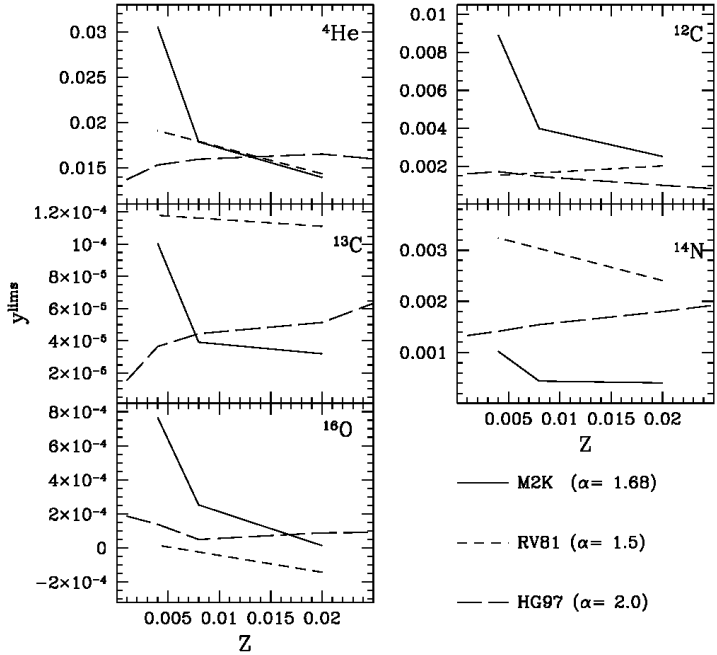}
\caption{Net yields of typical products of intermediate mass stars, integrated over an IMF and displayed as a function of initial stellar metallicity; data from various authors are compiled by Marigo (2001).}
\end{center}
\end{figure}

Some results of nucleosynthesis calculations for AGBs are displayed in Fig. 5 (integrated over an IMF) as a function of initial stellar metallicity. It is clearly seen that $^{14}$N and $^{13}$C are produced as primaries (because of Hot Bottom Burning), whereas $^{16}$O is depleted (net yields are negative). Yields of s-nuclides in AGBs as a function of metallicity are given in Goriely and Mowlavi (2000).

\subsubsection{Yields of thermonuclear supernovae (SNIa)}

Thermonuclear supernovae are exploding white dwarfs in binary systems (see Isern, this volume). Although neither the progenitor stars nor the exact mode of the propagation of the nuclear flame are well understood at present, it is generally believed that a large fraction of the white dwarf mass burns in conditions of Nuclear Statistical Equilibrium (NSE). In those conditions, large amounts of Fe-peak nuclei and, in particular, $^{56}$Ni, are produced: about 0.6 \ms \ of $^{56}$Ni, with extreme values ranging from up to 1 \ms \ down to 0.1 \ms. 

In most models, only the internal parts of the white dwarf burn all the way to NSE; intermediate mass nuclei (from Si to Ca) are also produced in the explosion, in agreement with observations of early spectra of SNIa. The most widely used model is the deflagration model W7, which reproduces satisfactorily the spectroscopic requirements. The results of a recent version of that model (Iwamoto et al. 1999) appear in Fig. 6. One may note a substantial production of Si, S, Ar and Ca (at about 30\% of the corresponding Fe production) and a slight oveproduction of $^{54}$Fe and $^{58}$Ni (the major default of the W7 model on grounds of nucleosynthesis).

\begin{figure}
\begin{center}
 \includegraphics[width=6cm,angle=-90]{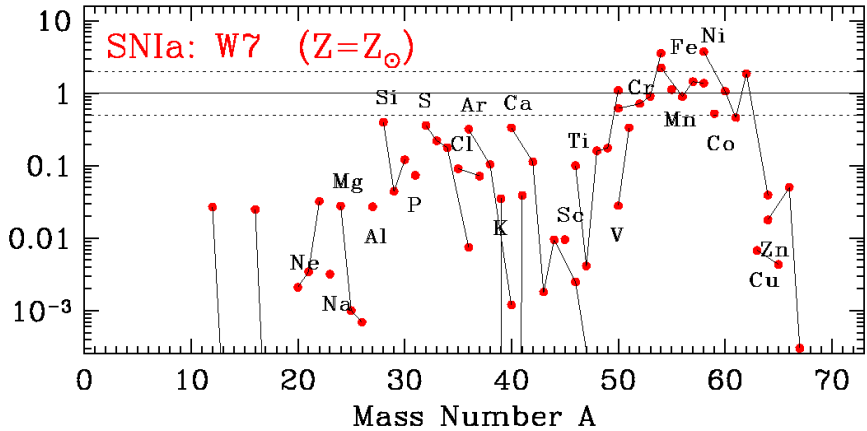}
\caption {Production factors $f$ for isotopes from C to Cu in the ``canonical'' model W7 of SNIa (from Iwamoto et al. 1999). They are normalized to $f_{Fe56}$=1.}
\end{center}
\end{figure}

SNIa are major contributors to the galactic content of $^{56}$Fe, the stable product of $^{56}$Ni decay, and in Fe-peak nuclides in general. In the case of the Milky Way, this can be seen as follows: The observed frequency of SNIa in external galaxies of the same morphological type (i.e. Sbc/d) is about 5 times smaller than the corresponding frequency of core collapse SN (SNII+SNIb,c), see e.g. Mannucci et al. (2005). But core collapse SN produce, on average $\sim$0.1 \ms \ of $^{56}$Ni, that is 6 times less than SNIa. Thus, SNIa contribute at least as much as massive stars to the production of Fe and Fe-peak nuclides in the Milky Way. In the case of the solar neighborhod, this is coroborated by another observational argument, namely the evolution of the O/Fe ratio,  that
will be presented in Sec. 3.1.2.

SNIa may have long-lived progenitors, with lifetimes of up to several Gyr; this introduces a substantial delay in their rate of element ejection in the ISM. The evolution of SNIa rate depends on the assumptions made about the progenitor system and it obviously cannot be simply proportional to the SFR. 
One of the simplest parametrizations, introduced recently,  consists in assuming  that the SNIa rate is proportional to both the SFR {\it and} the cumulative mass of the stellar population
\begin{equation}
R_{SNIa}(t) \ = \ \alpha \Psi(t) \ + \beta \ m_S(t)
\end{equation}
where $\Psi(t)$ and $m_S(t)$ are defined in Eq. (2.2) and (2.4), respectively. 
The values of $\alpha$ and $\beta$ are adjusted, as to account  for the observations of SN rates in external galaxies of different morphological types. It should be noted, however, that other formulations provide an equally satisfactory fit to the data. This is the case of the widely used parametrisation of Greggio and Renzini (1983), taking into account the time delay introduced by the evolution of the secondary star and the mass distribution of the two components of the binary system. Despite their uncertainties, due to our poor understanding of the progenitor systems of SNIa, such formulations offer more physical insight into the problem.

\subsection{Initial Mass Function (IMF)}

The IMF cannot be calculated at present from first principles (despite a large body of theoretical work, see e.g. Bonnell et al. 2006 for a recent review) and it has to be derived from observations. However, this derivation is not quite straightforward and important uncertainties still remain, especially in the region of massive stars.

In a first step, from the observed {\it Luminosity function} $f(L)=dN/dL$ and the theoretical mass-luminosity relation for Main sequence stars $\phi(L)=dM/dL$ , the {\it Present Day Mass Function} $F(M)$ is derived 
\begin{equation}
F(M) \ = \ {{dN}\over{dL}} \ {{dL}\over{dM}}
\end{equation}
Assuming that $\phi(L)$ is well known, one has still to account for various biases in the derivation of $F(M)$, like e.g. the role of binary and multiple stellar systems. In the case of stellar clusters, the impact of cluster dynamics (leading to the preferential evaporation of the lowest mass members) has to be taken into account. In the case of the field $F(M)$ in the solar neighborhood, one has to account for the fact that the dispersion of stellar velocities perpendicularly to the galactic plane increases with time, i.e. the corresponding scale-height of the stellar population $h(L)$ is a decreasing function of $L$.

Once the bias-corrected PDMF is established, the IMF may be derived, in principle, {\it if 
the Star formation rate (SFR) is known}. In fact, the PDMF is the integral over time of the {\it star creation rate } $C(M,t)=\Phi(M) \Psi(t)$, from which dead stars are removed. Assuming that the IMF deos not depend on time, one has:
\begin{equation}
\Phi(M) \ = \ {{F(M)}\over{\int^T_{T-\tau_M} \Psi(t) dt}}
\end{equation}
where $T$ is the age of the system ($T\sim$12-13 Gyr in the case of the solar neighborhood and the Milky Way disk). That expression can be simplified in two cases, where knowledge of $\Psi(t)$ is not really required:
\begin{itemize}
\item{In the case of short-lived stars ($\tau_M<<T$, i.e. $M>2$ \ms)
\begin{equation}
\Phi(M) \ = \ {{F(M)}\over{\Psi_0 \ \tau_M}}
\end{equation}
where $\Psi_0$ is the current SFR, assumed to be constant during the last Gyr.}
\item{In the case of ``eternal'' stars ($\tau_M>>T$, i.e. $M<0.8$ \ms \ for low metallicity stars)
\begin{equation}
\Phi(M) \ = \ {{F(M)}\over{<\Psi> \ \tau_M}}
\end{equation}
where $<\Psi>=\int^T_0 \Psi(t) dt/T$ is the {\it past average SFR}
during the system's history.}
\end{itemize}
Note that the {\it shape} of the IMF $\Phi(M)$ in Eq. (2.21) or (2.22) does not depend on the values of  $\Psi_0$ or $<\Psi>$. For the intermediate mass regime (0.8$< M/M_{\odot}<$2) the situation is more complicated and knowledge of $\Psi(t)$ is, in principle, required. Continuity arguments may be used to bridge the gap between the low and high mass regimes. The evolution of the $PDMF$ in the solar neighborhood is illustrated in Fig. 8 (left panel), where it is also compared to field star observations (right panel).

\begin{figure}
\begin{center}
 \includegraphics[width=8cm,angle=-90]{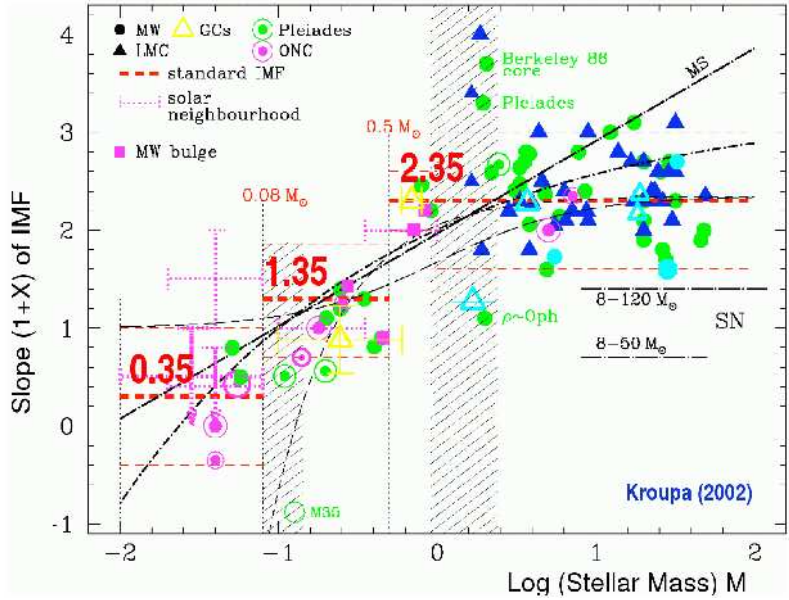}
\caption{Slope $1+X$ of the IMF (assumed to be described by a multi-power-law form), according to observations in various asrophysical environments; the dashed horizontal lines indicate average values in three selected mass ranges, with $1+X$=2.35 being the classical Salpeter value (from Kroupa 2002).}
\end{center}
\end{figure}

\begin{figure}
\begin{center}
 \includegraphics[width=6cm,angle=-90]{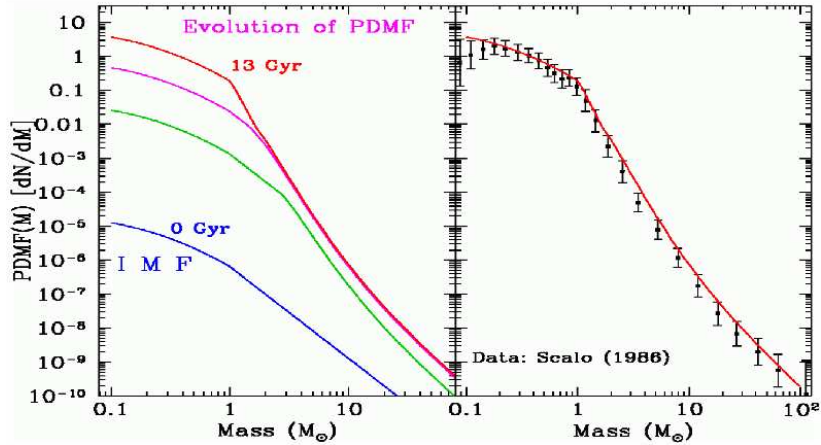}
\caption{{\it Left:} Evolution of the Present Day Mass Function, displayed here at four different moments (0, 1, 4 and 13 Gyr) for a smooth star formation history, correspondig to the one of the solar neighborhood. The initial PDMF (0 Gyr) is the same as the adopted IMF, while the final PDMF (13 Gyr) is the one observed today. The latter curve is compared on the {\it right} panel with the PDMF derived for the solar neighborhood by Scalo (1986, {\it error bars}). The adopted IMF is log-normal below 1 \ms \ and a power-law with slope $X$=1.7 above 1 \ms.}
\end{center}
\end{figure}

Based on observations of stars in the solar neighborhood and accounting for various biases (but not for stellar multiplicity), Salpeter (1955) derived the local IMF in the mass range 0.3-10 \ms, a power-law IMF:
\begin{equation}
\Phi(M) \ = \ {{dN}\over{dM}} \ = \ A \ M^{-(1+X)}
\end{equation}
with slope $X$=1.35. That slope is indeed found in a large variety of conditions and the Salpeter IMF is often used in the whole stellar mass range, from 0.1 to 100 \ms, especially in studies of the photometric evolution of galaxies. However, it is clear now that there are less stars in the low mass range (below 0.5 \ms) than predicted by the Salpeter slope of $X$=1.35. For instance, Kroupa (2002 and Fig. 7) adopts a multi-slope power-law IMF with $X$=0.35 in the range 0.08
to 0.5 \ms, whereas Chabrier (2003) prefers a log-normal IMF below 1 \ms.

\begin{figure}
\begin{center}
 \includegraphics[width=8cm,angle=-90]{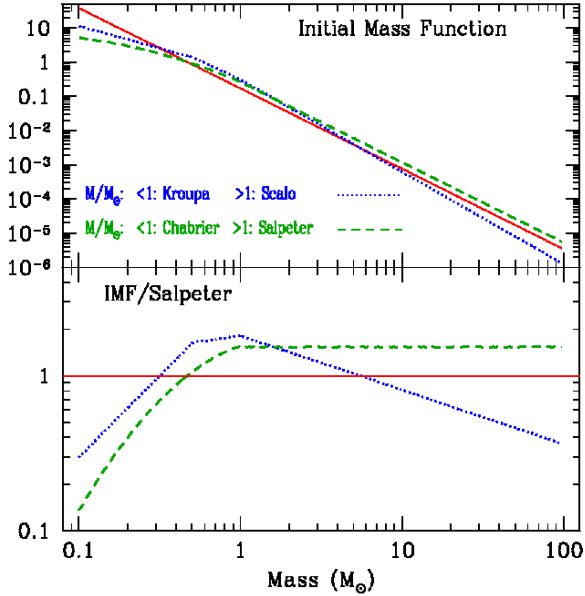}
\caption{ {\it Top:} Three initial mass functions: {\it solid curve}: Salpeter (power-law in the whole mass range), {\it dotted curve}: Kroupa (multi-slope power law for M$<$1 \ms) + Scalo ($X$=1.7 for M$>$1 \ms), {\it dashed curve}: Chabrier (log-normal  for M$<$1 \ms) + Salpeter ($X$=1.35 for M$>$1 \ms). {\it Bottom:} Ratio of the three IMFs to the one of Salpeter.}
\end{center}
\end{figure}

For GCE purposes, low mass stars are ``eternal'' and just block matter from recycling in the ISM. Most important, in that respect, is the shape of the IMF above 1 \ms. Unfortunately, the situation is not  clear yet. Observations of the IMF in various environments, and in particular young clusters (where dynamical effects are negligible) suggest that a Salpeter slope $X$=1.35 describes the data well (see also Fig. 7). However, determination of the IMF in young clusters suffers from considerable biases introduced by stellar multiplicity and pre-main sequence evolution. Moreover, in the case of the field star IMF in the the solar neighborhood, Scalo (1986) finds $X$=1.7, i.e. a much steeper IMF than Salpeter.

According to Weidner and Kroupa (2006), the slope of the {\it stellar IMF} is indeed the one observed in clusters, but GCE studies involve the  {\it galaxian IMF}, i.e. the sum of all cluster IMFs, which is steeper than the stellar IMF. The reason is that, although every single cluster has the same stellar IMF  ($X$=1.35), the maximum stellar mass $M_{MAX,C}$ in a  a cluster increases with the total mass of that cluster: indeed, observations reveal that small clusters may have  $M_{MAX,C}$ as low as a few \ms, whereas large clusters have  $M_{MAX,C}$ up to 150 \ms. If this were just a statistical effect, the slope of the resulting galaxian IMF (i.e. the sum over all cluster IMFs) would also be $X$=1.35, but if there is a physical reason for the observed $M_{MAX,C}$ vs $M_{Cluster}$ relation, then the  resulting galaxian IMF would necessarily be steeper. In that way, one may conciliate the results of Scalo (1986, a study still unique in its kind) with observations of cluster IMFs, which favour the Salpeter value\footnote{Some implications of that idea, suggesting that the slope of the high mass part of the IMF is variable and depends on the star forming activity of the galaxy, are explored in K\"oppen et al. (2006).}. For GCE studies, the Scalo value of $X$=1.7 appears more appropriate, and it is indeed used in most detailed models of the local GCE.

The IMF is normalised to
\begin{equation}
\Phi(M) \ = \ \int^{M_U}_{M_L} \ \Phi(M) \ M \ dM \ = \ 1
\end{equation}
where $M_U$ is the upper mass limit and $M_L$ the lower mass limit. Typical values are $M_U\sim$100 \ms \ and  $M_L\sim$0.1 \ms, and the results depend little on the exact values (if they remain in the vicinity of the typical ones). A comparison between three normalized IMFs, namely the ``reference'' one of Salpeter, one proposed by Kroupa (with the Scalo slope at high masses) and one by Chabrier (with the Salpeter slope at high masses) is made in Fig. 9.

A useful quantity is the {\it Return Mass Fraction} $R$ 
\begin{equation}
R \ = \ \int^{M_U}_{M_T} \ (M-C_M) \ \Phi(M) \ dM
\end{equation}
i.e. the fraction of the mass of a stellar generation that returns to the ISM. For the three IMFs displayed in Fig. 9 one has $R$= 0.28 (Salpeter), 0.30 (Kroupa+Scalo) and 0.34 (Chabrier+Salpeter), respectively, i.e. about 30\% of the mass gone into stars returns to the ISM.

\subsection{Star Formation Rate}

Star formation is the main driver of galactic evolution and  the most uncertain parameter in GCE studies. Despite decades of intense observational and theoretical investigation (see e.g. Elmegreen 2002 and references therein) our understanding of the subject remains frustratingly poor.

Observations of various SFR tracers in galaxies provide only relative values, under the assumption that the IMF is the same everywhere. Most such tracers concern formation of stars more massive than $\sim$2 \ms; very little information exists for the SFR of low mass stars, even in the Milky Way.

Some commonly used tracers of SFR (i.e. of young stellar populations) are:
\begin{itemize}
\item{Luminous star counting in some region of the Hertzsprung-Russell diagram; this method can be applied only to nearby galaxies like the Magellanic clouds.}
\item{Flux measurements in recombinations lines H$_{\alpha}$, H$_{\beta}$ etc., emitted by gas which is ionized by nearby OB stars (emitting photons bluewards of 912 A).}
\item{UV flux of OB stars, which are relatively young (the UV emissivity of Horizontal branch stars and of planetary nebulae nuclei is negligible, in general).}
\item{Far infrared (FIR) emission of dust, heated by absorption of UV photons emitted by hot nearby stars.}
\item{Surface density of supernova remnants, like pulsars (in star forming galaxies, i.e  spirals and irregulars).}
\end{itemize}

Those tracers reveal that star formation occurs in different ways, depending on the type of the galaxy. In spiral disks, star formation occurs mostly inside spiral arms, in a sporadic way. In dwarf, gas rich, galaxies, it occurs in a small number of bursts, separated by long intervals of inactivity. Luminous Infrared galaxies (LIRGS) and starbursts (as well as, most probably, ellipticals in their youth) are caracterised by an intense burst of star formation, induced by the interaction (or merging) with another galaxy.

Today there is no theory to predict large scale star formation in a galaxy, given the various physical ingredients that may affect the SFR (e.g. density and mass of gas and stars, temperature and composition of gas, magnetic fields and frequency of collision between giant molecular clouds, galactic rotation etc.)
Only in the case of spirals predictions can be made, based on various instability criteria (e.g. Elmegreen 2002 for a review). In the case of other galaxies, a parametrised formulation of the SFR is adopted.  Schmidt (1959) suggested that the density of the SFR  $\Psi$ is proportional to some power of the density of  gas mass $m_G$:
\begin{equation}
\Psi \ = \ \nu \ m_G^N
\end{equation}
a formulation which has the merit of reminding us that stars are formed from gas, after all. However, it is not clear whether {\it volume density} $\rho$ or {\it surface density} $\Sigma$ should be used in Eq. 2.26. When comparing data with GCE models for the solar neighborhood, Schmidt (1959) uses surface densities ($\Sigma$ in \ms/pc$^2$). But, when finding "direct evidence for the value of $N$" in his paper\footnote{Schmidt(1959) describes the distributions of gas and young stars perpendicularly to the galactic plane ($z$ direction) in terms of {\it volume densities} $\rho_{Gas} \propto exp(-z/h_{Gas})$ and $\rho_{Stars} \propto exp(-z/h_{Stars})$ with corresponding scaleheights (obervationally derived) $h_{Gas}$=78 pc and $h_{Stars}$=144 pc$\sim$2 $h_{Gas}$; from that, Schmidt deduces that $\rho_{Stars} \propto \rho_{Gas}^2$, that is $N$=2.} he uses volume densities ($\rho$ in \ms/pc$^3$) and finds $N$=2. Obviously, since $\Sigma$=$\int_z \rho(z) dz$, one has: $\Sigma^N \neq \int_z \rho(z)^N dz$.

\begin{figure}
\begin{center}
 \includegraphics[width=6cm,angle=-90]{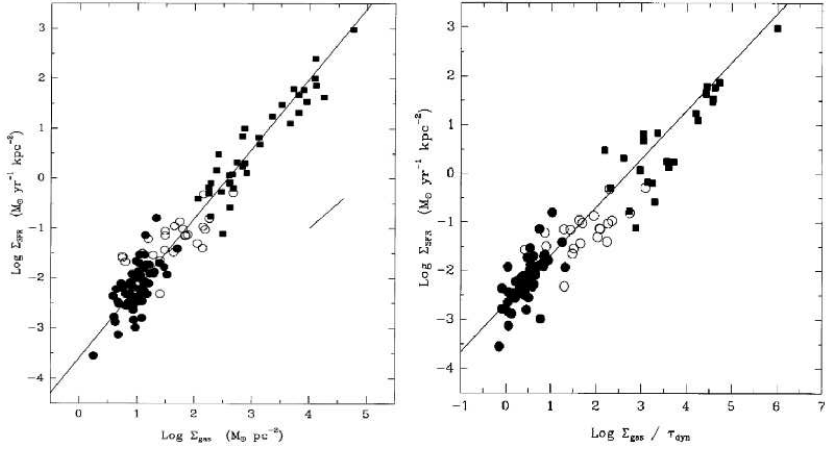}
\caption{{\it Left:} Average surface density of star formation rate $\Psi$ (in M$_{\odot}$ yr$^{-1}$ pc$^{-2}$) as a function of average gas (HI+H$_2$) surface density $\Sigma_G$ (in M$_{\odot}$ pc$^{-2}$, in spirals (circles) and starbursts (squares); the solid line corresponds to $\Psi \propto \Sigma^{1.4}$. {\it Right}:  Average surface density of star formation rate $\Psi$ (in M$_{\odot}$ yr$^{-1}$ pc$^{-2}$) as a function of $\Sigma_G/\tau_{dyn}$, where the dynamical timescal $\tau_{dyn} = R/V$ (for rotational velocity $V$ at radius $R$); the solid line corresponds to $\Psi \propto \Sigma_G/\tau_{dyn}$ (from Kennicut 1998).}
\end{center}
\end{figure}

It is not clear then which density should be used in the Schmidt SFR law: Volume density is more ``physical'' (denser regions collapse more easily) but surface density is more easily measured in  galaxies. Furthermore, at first sight, it seems that the density of molecular gas should be used (since stars are formed from molecular gas), and not the total gas density.

Surprisingly enough, Kennicutt (1998) finds that, in normal spirals, the surface density of SFR  correlates  with atomic rather than with molecular gas;
this conclusion is based on {\it average surface densities}, i.e. the total SFR and gas amounts of a galaxy are divided by the corresponding surface aerea of the disk. In fact, Kennicut (1998) finds that a fairly good correlation exists between SFR and {\it total (i.e. atomic + molecular) gas}. This correlations extends over four orders of magnitude in average gas surface density $\rho_S$ and over six orders of magnitude in average SFR surface density $\Psi$, from normal spirals to active galactic nuclei and starbursts (see Fig. 10, left) and can be described as:
\begin{equation}
\Psi \ \propto  \ \Sigma^{1.4}
\end{equation}
i.e. $N$=1.4. However, Kennicutt (1998) notes that the same data can be fitted equally well by a 
different $N$ value, this time involving the {\it dynamical timescale} $\tau_{dyn} \ = \ R/V(R)$, 
where $V(R)$ is the orbital velocity of the galaxy at the optical radius $R$:
\begin{equation}
\Psi \ \propto  \ {{\Sigma}\over{\tau_{dyn}}}
\end{equation}
More recent observations confirm Kennicutt's results, but are also compatible with other SFR laws suggested in the literature (see e.g. Boissier et al. 2003). Note that, for a given data set, the determination of coefficient $\nu$ in Eq. 2.26 depends on the determination of exponent $N$, and vice versa.

\begin{figure}
\begin{center}
 \includegraphics[width=8cm0]{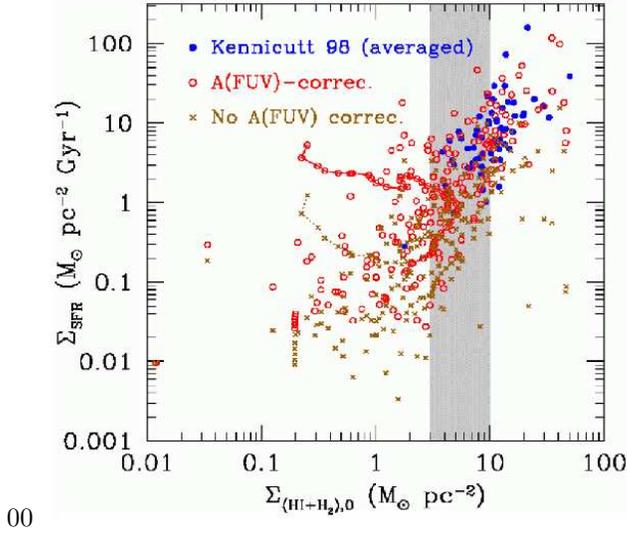}
\caption{Surface density of star formation rate $\Psi$   as a function of gas (HI+H$_2$) surface density $\Sigma_G$ in galaxies observed with GALEX. The vertical band corresponds to the SFR treshold suggested by Kennicutt (1998). Clearly, UV observations show SFR activity below that threshold (from Boissier et al. 2006).}
\end{center}
\end{figure}

In some cases, it is useful to consider the {\it efficiency of star formation $\varepsilon$}, i.e. the SFR per unit mass of gas. In the case of a Schmidt law with $N$=1 one has: $\varepsilon = \nu=const.$, whereas in the case of $N$=2 one has: $\varepsilon = \nu m_G$.

Finally, one may ask whether the SFR law in Fig. 10 extends down to lower surface densities or whether there is a {\it  threshold} of star formation in galactic disks. The data of Kennicutt (1998 and Fig. 10) apparently suggest a lower threshold of a few \ms/pc$^2$. However, more recent data, obtained with the GALEX UV satellite, reveal star formation activity even in regions of gas density as low as 0.1 \ms/pc$^2$ (Fig. 11). Those recent data imply that the threshold  of star formation is much lower than thought before (and, perhaps, non existant).

\subsection{Gaseous flows and stellar feedback}

A galaxy is clearly not an isolated system, and it is expected to exchange matter (and energy) with its environment. This is true even for galaxies which are found away from galaxy groups.
First of all, most of baryonic matter in the Universe today (and in past epochs) is in the form of gas  residing in the intergalactic medium (e.g. Fukugita and Peebles 2004) and  part of it is slowly accreted by galaxies. Also, small galaxies are often found in the tidal field of larger ones, and their tidal debris (gas and/or stars) may be captured by the latter.  In both cases, gaseous matter is accreted by galaxies, and in the framework of the simple GCE model this is generically called {\it infall}.

On the other hand, gas may leave the galaxy, if it gets  sufficient (1) kinetic energy or (2) thermal energy and (3) its velocity becomes larger than the escape velocity. Condition (1) may be met in the case of tidal stripping of gas in the field of a neighbor galaxy or in the case of ram pressure from the intergalactic medium. Condition (2) is provided by heating of the intestellar gas from the energy of supernova explosions, especiallly if collective effects (i.e. a large number of SN in a small volume, leading to a superbubble) become important. Finally, condition (3) is more easily met in the case of small galaxies, with swallow potential wells. Note that, since galaxies (i.e. baryons) are embedded in  extended dark matter (non baryonic)  haloes, a distinction should be made between gas leaving the galaxy but still remaining trapped in the dark halo and gas leaving  even the dark halo. In the former case, gas may return back to the galaxy after ``floating'' for some time in the dark halo and suffering sufficient cooling. In the framework of the simple GCE model, all those cases are described generically as {\it outflows}.

The rate of infall or outflow is difficult to calculate from first principles. In the case of infall, this is possible, in principle, for a hydrodynamical model evolving in an appropriate cosmological framework. In the case of outflows, the interaction between stars (SN) and the ISM, known as {\it feedback}, also requires detailed hydrodynamic modelling. No satisfactory models exist up to now for such complex processes. Incidentally, the treatment of feedback also affects the SFR of the system (by making gas unavailable for star formation, either by heating it or by pushing it out of the system alltogether).

In simple GCE models, infall and outflow are treated as free parameters, adjusted as to reproduce  observed features of the galaxian systems under study. Such features are the metallicity distributions (MD) of long -lived stars, or the mass-metallicity relationship of external galaxies. The former feature is nicely illustrated in the cases of the MD in the local disk and the  halo (Sec. 3.1 and 3.2, respectively), which provide strong constraints on the history of those systems.

\subsection{Analytical solutions: the Instantaneous Recycling Approximation (IRA)}

The system of GCE equations (2.1 to 2.9) can be solved analytically if one adopts the Instantaneous Recycling Approximation (IRA), introduced by Schmidt (1963). Stars are divided in: ``eternal'' ones (those of low mass, with lifetimes far exceeding the age of the system) and ``dead at birth'' ones (massive stars, with lifetimes far shorter than the age of the system) for which it is assumed that $\tau_M$=0. The dividing line between the two classes depends on the system's age, and for most practical purposes it is put at 1 \ms, corresponding to an age $T\sim$12 Gyr.

Assuming IRA allows one to replace $\Psi(t-\tau_M)$ in the equations  of GCE by $\Psi(t)$ and thus to take out the SFR $\Psi(t)$ from the mass integrals. The equations can then be solved analytically (see e.g. Tinsley 1980). Solutions involve the {\it Return fraction R} defined in Eq. 2.25 and the {yield} $p_i$ of a given nuclide, defined as:
\begin{equation}
p_i \ = \  {{1}\over{1-R}} \ \int^{M_U}_{M_T} \ y_i(M) \ \Phi(M) \ dM
\end{equation}
The yield $p_i$ is {\it the newly created amount of nuclide i by a stellar generation, per unit mass of stars blocked into ``eternal''  objects}: indeed, $y_i(M)$ are the {\it net yields} of nuclide $i$ (see Sec. 2.3.1) and $1-R$ is the mass blocked in low mass stars and compact objects (for the normalized IMF of Eq. 2.24). Obviously, the {\it stellar yields} $Y_i(M)$ and {\it net yields} $y_i(M)$ are properties of individual stars, while the {\it yield} $p_i$ is an integrated property of the IMF, as is the return fraction $R$. Note that only $Y_i(M)$ are interesting for numerical models, whereas $y_i(M)$ and $p_i$ are useful only in analytical models.

\begin{figure}
\begin{center}
 \includegraphics[width=6cm,angle=-90]{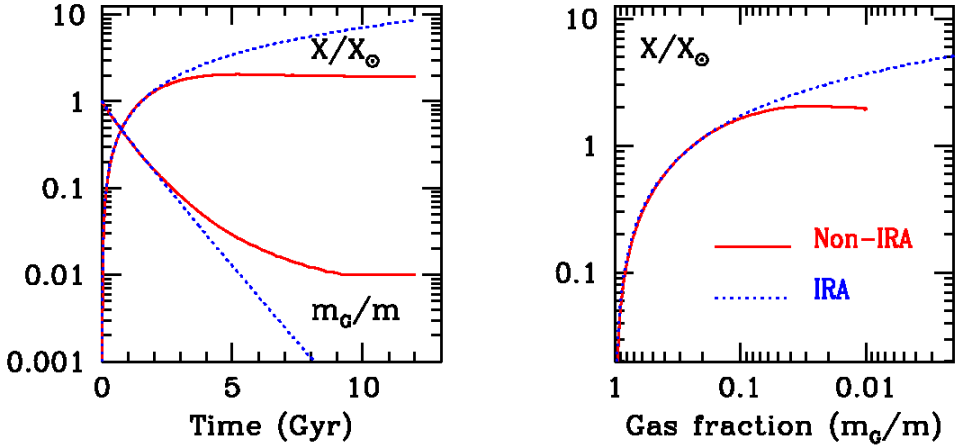}
\caption{Results of calculations for a Closed Box model with Instantaneous Recycling Approximation (IRA, {\it dotted curves}) vs Non-IRA ({\it solid curves}). {\it Left:} Metallicity (upper curves) and gas fraction (lower curves) as a function of time. {\it Right:} Metallicity as a function of gas fraction. It is assumed that $\Psi = 1.2 \ m_G$ Gyr$^{-1}$.}
\end{center}
\end{figure}

With those definitions, the gas mass fraction $X_i$ of nuclide $i$  in the case of the Closed Box model is:
\begin{equation}
X_i \ - \ X_{i,0} \ = \ p_i \ ln({{m}\over{m_G}}) \ = \ p_i \ ln({{1}\over{\sigma}})
\end{equation}
where $\sigma=m_G/m$ is the {\it gas fraction} and $X_{i,0}$ the initial abundance ($X_{i,0}$=0 for metals). This is the main result of IRA, relating the chemical enrichment of the gas to the amount of gas left. 
It is independent on time or on the form of the SFR, and  for those reasons it is a powerful tool to study gas flows in the system. It can also be used to derive
the metallicity distribution of stars and, thereof, the past history of the system (see Sec. 3.1.1 and 3.2.2).

If a simple Schmidt law of the form $\Psi=\nu m_G$ is adopted for the SFR (Eq. 2.26) one can obtain the following solutions for the evolution of gas
\begin{equation}
m_G \ = \ m \ e^{-\nu (1-R)t}
\end{equation}
and for the abundances
\begin{equation}
 X_i \ - \ X_{i,0} \ = \ p_i \ \nu \ \ (1-R) \ t
 \end{equation}
 which also satisfy the more general solution of Eq. 2.30. Thus, metallicity is roughly proportional to  time, a result which is approximately valid even when IRA is relaxed. This property allows one to use stellar metallicity (especially iron, which has many strong and easily identifiable spectral lines) as a proxy for time, since stellar ages are notoriously difficult to evaluate.

IRA turns out to be a surprisingly good aproximation for nuclides produced in massive stars, like e.g. oxygen, provided gas fractions stay above $\sim$10\%; this is illustrated in Fig. 12 for the case of a Closed Box model, with analytical solutions given by 2.31 and 2.32. Analytical solutions assuming IRA can also be obtained in the cases of gaseous flows into or out of the system. However, some specific assumptions have to be made about the form of those flows, and this limits a lot their interest. For instance, in the case of outflow at a rate proportional to the SFR, $o \ = k \ \Psi$, metalicity evolves as:
\begin{equation}
 X_i \ - \ X_{i,0} \ = \ {{p_i}\over{1+k}} \ ln({{1}\over{\sigma}})
\end{equation}
i.e. at the same gas fraction metallicity is smaller than in the Closed box or, equivalently, a larger fraction of the system has to turn into stars in order to reach the same metallicity. Eq. 2.33 is formally the same as 2.30, with an {\it effective yield}
\begin{equation}
p_{i,eff} \ = \  {{p_i}\over{1+k}}
\end{equation}
in the place of the true yield $p_i$, and $p_{i,eff} < p_i$. It is shown (Edmunds 1990) that gas flows always produce {\it reduced effective yields}, i.e. that metallicity increases most efficiently in the Closed Box model. For other analytical solutions, especially for models with specific forms of infall, see e.g. Pagel (1997) or Matteucci (2001).

In the case of a secondary nuclide (see Sec. 2.3.2), its yield $p_S$ is proportional to the abundance of a primary one, i.e. $p_S \ = \ \alpha \ X_P$, and its evolution in the framework of Closed Box plus IRA is described by:
\begin{equation}
X_S \ = \  \alpha \ X_P \ ln({{1}\over{\sigma}}) \ = \ {{\alpha}\over{p_P}} \ X_P^2
\end{equation}
that is, its abundance increases faster than the abundance of the primary ($p_P$ being the yield of the primary).

The abundance ratio of two primaries or two secondaries remains constant during galactic evolution and equal to the ratio of the corresponding yields, independently of the IMF and the SFR. Observations of that ratio allow then one to test theories of stellar nucleosynthesis, independently of the underlying GCE model. This is true only if both nuclides are produced in the same site (e.g. silicon and calcium, products of massive stars). On the other hand, an element produced mainly in a long-lived site  may appear late in the Galaxy, even if it is a primary one. This is, among others, the case of Fe: its observed abundance increases faster than the one of oxygen after the first Gyr, and this increase is attributed not to a secondary nature but to a delayed production in SNIa (see Sec. 3.1.2).

\section{The Chemical evolution of  the Solar Cylinder}

Obviously, the larger the number of observables, the more the framework provided by GCE can be useful in constraining the history of the system. The Milky Way and, in particular, the Solar cylinder,  is the best observed system today (for obvious reasons).

The Solar cylinder may be  defined as a cylindrical region of radius 0.5 kpc, perpendicular to the galactic plane and  centered at Sun's position (at 8 kpc from the Galactic center, see Fig. 13). The interstellar gas is located near the plane and it is (usually assumed to be) chemically homogeneous.  On the other hand,  three stellar populations coexist, to various extents, in that region; they are distinguished by their kinematic and chemical properties.
\begin{itemize}
\item{ The {\it galactic halo}: old (age $\sim$12-13 Gyr), ``pressure'' supported, metal poor ([Fe/H]$<$-1) with little or no rotation, large velocity dispersion perpendicularly to the disk and small contribution ($<$5\%) to the overall surface density.}
\item{ The {\it thin disk}: young (age $\sim$5-6 Gyr on average), rotationally supported, metal rich ([Fe/H]$\sim$-0.1 on average), with small scaleheight ($h\sim$300 pc for the old stars, but only $h\sim$100 pc for the young stars and the gaseous layer), dominating the total mass (75\%).}
\item{The {\it thick disk}: older than the thin disk ($\sim$10 Gyr), more extended (scaleheight $h\sim$1.4 kpc), rotating, with moderate velocity dispersion, moderately metal poor ([Fe/H]$\sim$-0.7) and contributing $\sim$20\% to the total surface density.}
\end{itemize}
It is tempting to assume, in the framework of the old, {\it monolithic collapse} scenario for the Milky Way's evolution (Eggen et al. 1962), a temporal continuity
 in the formation of those three components, i.e.
halo $\longrightarrow$ thick disk $\longrightarrow$ thin disk\footnote{In fact, it is difficult to account for the chemical and kinematical properties of all three galactic components in the monolithic collapse scenario.}. Such a continuity is more difficult to establish in the modern framework of hierarchical galaxy formation: the three components may be totally uncorrelated, e.g. the thick disk may have been formed from tidal debris of satellite galaxies, while the thin disk from a slow accretion processes\footnote{The monolithic collapse scenario may describe better the properties of the inner halo and of the bulge.}. In the following we shall study the cases of the local thin disk (often called the Solar Neighborhood) and of the Milky Way's halo independently.

\begin{figure}
\begin{center}
 \includegraphics[width=9cm,angle=-90]{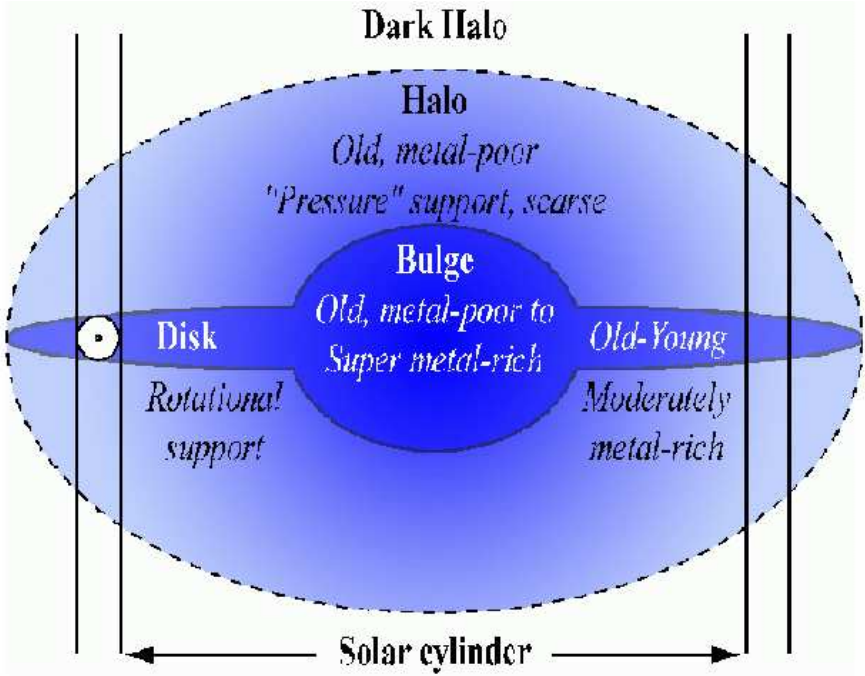}
\caption{The various components of the Milky Way (bulge, halo, disk) and their main features; a distinction should be mde between thin and thick disk (not appearing in the figure). The Solar cylinder is at 8 kpc from the center (adapted from Pagel 1997).}
\end{center}
\end{figure}

\subsection{The local thin disk}

In the case of the Solar neighborhood, the number of available observational data is larger than for any other galactic system, allowing one to constrain strongly (albeit not in a unique way)  the history of the system. Those data are
 (see e.g. Boissier and Prantzos 1999):
\begin{itemize}
\item{The current surface densities of gas ($\Sigma_G\sim$12 \mp), live stars (($\Sigma_*\sim$30 \mp), stellar residues ($\Sigma_C\sim$8 \mp) and total amount of baryonic matter ($\Sigma_T\sim$50 \mp), as well as the current star formation rate ($\Psi_0$=2-5 \mp \ Gyr$^{-1}$); the corresponding  gas fraction is $\sigma\sim$0.24.}
\item{The elemental and isotopic abundances at solar birth ($X_{i,\odot}$, i.e. it is assumed that the Sun's composition is typical of the ISM 4.5 Gyr ago) and  today ($X_{i,Now}$). Those two compositions are quite similar, suggesting little chemical evolution in the past 4.5 Gyr.}
\item{The stellar age-metallicity relationship (AMR), traced by the Fe abundance of long-lived stars, [Fe/H]=$f(t)$ or $log(Z)=f(t)$.}
\item{The metallicity distribution (MD) of long-lived G-type stars ${{dn}\over{d[Fe/H]}}$ or ${{dn}\over{d(logZ)}}$, showing that few of them were formed at [Fe/H]$<$-0.7 (0.2 $X_{Fe,\odot}$).}
\item{The oxygen vs. iron (O-Fe) relationship, interpreted in terms of a delayed (after $\sim$1 Gyr) enrichment of the ISM with products of SNIa, contributing to more than half of the solar Fe.}
\end{itemize}

 Among those constraints, the age-metallicity relationship  and the G-dwarf metallicity distribution are the most important ones. In principle, by combining them, one may derive straightforwardly the SFR history ($dn/dt$) of the solar neighborhood through:
\begin{equation}
{{dn}\over{dt}} \ = \ {{dn}\over{d(logZ)}} \ {{d(logZ)}\over{dt}}
\end{equation}
In practice, however, this is impossible, because of the sensitivity of the result to the slope of the adopted AMR: a small variation in the form of the AMR produces a dramatic effect on the resulting SFR history (see Prantzos 1997). For that reason, the local SFR history is reconstructed only indirectly: models of the GCE of the Solar neighborhood are developed, which must  satisfy all the aformentioned observational constraints. Among those constraints, the metallicity distribution is probably the most significant.

\subsubsection{The Local Metallicity Distribution of long-lived stars}

In the framework of the Closed box model of GCE with IRA, the MD of long-lived stars can be derived as follows. In order to reach a given metallicity $Z$, a certain amount of stars has to be created, given by (Eq. 2.4): $m_S=m-m_G$, or (by normalising to the total mass $m$):
\begin{equation}
n \ = \ 1 \ - \sigma
\end{equation}
where $\sigma=m_G/m$ is the gas fraction and $n=m_S/m$ the star fraction.  For  a system with a final metallicity $Z_1$ and star fraction $n_1=1-\sigma_1$, the {\it cumulative metallicity distribution} (CMD), i.e. the number of stars with metallicity
lower than $Z$ as a function of $Z$, is given by
\begin{equation}
{{n(<Z)}\over{n_1}} \ = \ {{1-\sigma}\over{1-\sigma_1}}
\end{equation}
By using the fundamental result of IRA, namely $\sigma=exp(-Z/p)$ (Eq. 2.30) and taking the derivatives of Eq. 3.3, one obtains the {\it differential metallicity distribution} (DMD):
\begin{equation}
{{d(n/n_1)}\over{d(log Z)}} \ = \ {{ln(10)}\over{1-exp(-Z_1/p)}} \ \ {{Z}\over{p}} \ e^{-Z/p}
\end{equation}
i.e. the number of stars per logarithmic metallicity interval as a function of metallicity $Z$. This relation has a maximum for $Z=p$, i.e. when the metallicity is equal to the yield (both metallicity and yield can be expressed in units of the solar abundance of the corresponding element). 

Eq. 3.4 for the Closed box model with initial metallicity $Z_0$=0 appears in Fig. 14, where it is compared to data for the local disk. It predicts many more stars at low metallicities than observed, a problem known as the "G-dwarf problem"\footnote{G-type stars are bright enough for a reasonably complete sample to be constructed and long-lived enough to survive since the earliest days of the disk; the same problem is encountered if F- or K- type stars are used.}.

\begin{figure}
\begin{center}
 \includegraphics[width=6cm,angle=-90]{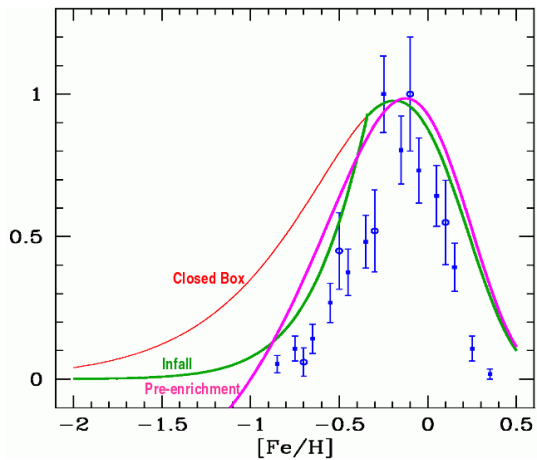}
\caption{Metallicity distribution of long-lived stars for three models: Closed Box, exponentially decreasing Infall (with a timescale of 7 Gyr) and Pre-enrichment (with $X_0 = 0.08 X_{\odot}$ for Fe), respectively. They are compared to data for the Solar neighborhood.}
\end{center}
\end{figure}

Two of the main solutions proposed for the G-dwarf problem appear also on Fig. 14. According to the first one, the disk started with an initial metallicity $Z_0\sim$0.1 \zs ({\it pre-enrichment}). In that case, all metallicities in Eq. 3.4 are replaced by $Z-Z_0$ and the resulting curve fits relatively well the data. The main drawback of that hypothesis is that it is hard to justify the origin of such a large pre-enrichment: it is true that the Galactic halo, which preceded disk formation, reached a maximum metallicity of $\sim$0.1 \zs \ (for Fe); but its average (stellar) metallicity is $\sim$0.03 \zs (for Fe) amd its total mass ($\sim$2 10$^9$ \ms) is 20 times smaller than the one of the disk ($\sim$4.5 10$^{10}$ \ms). There is simply not enough mass and metals produced in the halo to justify pre-enrichment of the disk to such a high level. Moreover, the halo has a low specific angular momentum (contrary to the disk) and material escaping it should be accreted rather by the bulge, not the disk.

The second hypothesis is that the disk did not evolve as a Closed box, but was gradually built from {\it infall} of metal free (or metal poor) material. In the Closed box, all the gas of the system is available from the very beginning; a large stellar activity is then required to enrich all that gas to, say, 0.1 \zs, and correspondingly many small and long-lived stars are formed at low $Z$. In the case of infall, only a small amount of gas exists early on; it takes then a small number of SN to enrich it to 0.1 \zs, and correspondingly  few low mass stars are formed at low $Z$. 

Infall appears then as an elegant solution to the local G-dwarf problem, especially in view of the fact that gas accretion to galaxies is  expected to be a common phenomenon in the Universe. The rate of the infall is not precisely determined by the data of the local disk. An exponentially decreasing infall rate $f(t)=A \ exp(-t/\tau)$ with a long  characteristic time-scale of $\tau\sim$7 Gyr provides a reasonably good fit to the data and, in view of its simplicity, is often used in models of the solar neighborhood (see next section). However, other forms may do as well and even better; this is the case, for instance, of a gaussian (as a function of time) infall rate, with a maximum prior to solar system formation (see e.g. Prantzos and Silk 1998). 

The infall rate $f(t)$ must obviously  be normalized
such as:
\begin{equation}
\int_0^T f(t) dt \ = \ \Sigma_T
\end{equation}
where $\Sigma_T$ is the total surface density in the Solar neighborhood (see  Sec. 3.1) and $T\sim$12 Gyr is the age of the Galactic disk.

\begin{figure}
\begin{center}
 \includegraphics[width=9cm,angle=-90]{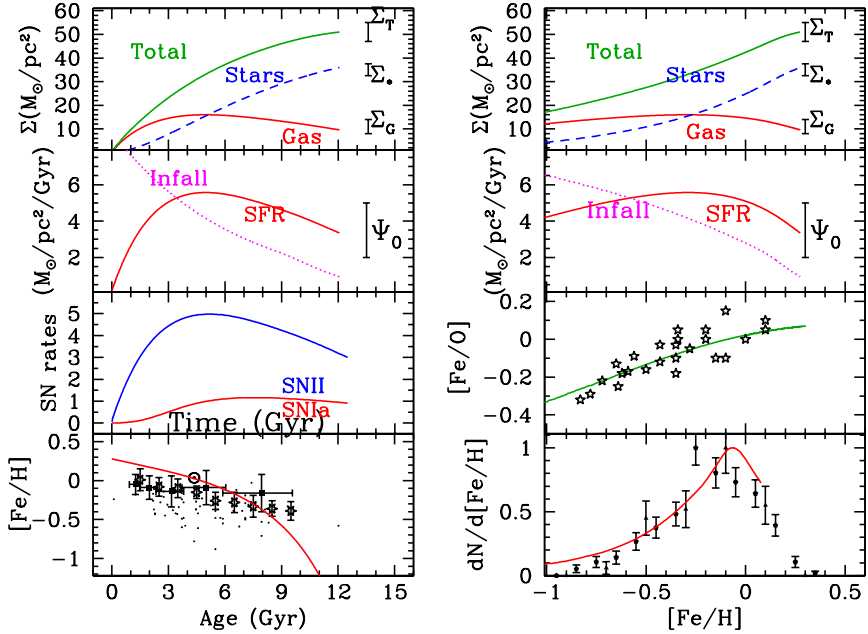}
\caption{History of the Solar neighborhood, according to a GCE model with infall, constrained by various observables.
{\it Left:} Results are plotted as a function of time (or age, for the bottom panel). Data for the present-day local disk are displayed with vertical bars. Data for the age-metallicity relation are from various sources. {\it Right:} Results are plotted as a function of metallicity [Fe/H]. See text for comments on the various curves.}
\end{center}
\end{figure}

\subsubsection{A brief History of the Solar neighborhood}

The observed properties of the local disk, presented in Sec. 3.1, along with the constraint of G-dwarf MD (analysed in Sec. 3.1.1), ``dictate'' the parameters of simple GCE models that may be built for that system. 
The results of such a model are displayed in Fig. 15, as a function of time (left panels) and of metallicity [Fe/H] (right panels). The various paremeters of the model are adjusted as follows:
\begin{itemize}
\item{The total amount of infalling matter is normalized to the local surface density through Eq. 3.5.}
\item{The time-scale  of the adopted exponentially decreasing infall rate is sufficiently long ($\tau$= 7 Gyr)  as to reproduce the metallicity distribution.}
\item{The coefficient $\nu$=0.3 Gyr$^{-1}$  of the SFR $\Psi = \nu \ m_G$ (Eq. 2.26) is adjusted as to leave the system at $T$=12 Gyr with a gas fraction $\sigma\sim$0.2, as observed.}
\end{itemize}

\begin{figure}
\begin{center}
 \includegraphics[width=9cm,angle=-90]{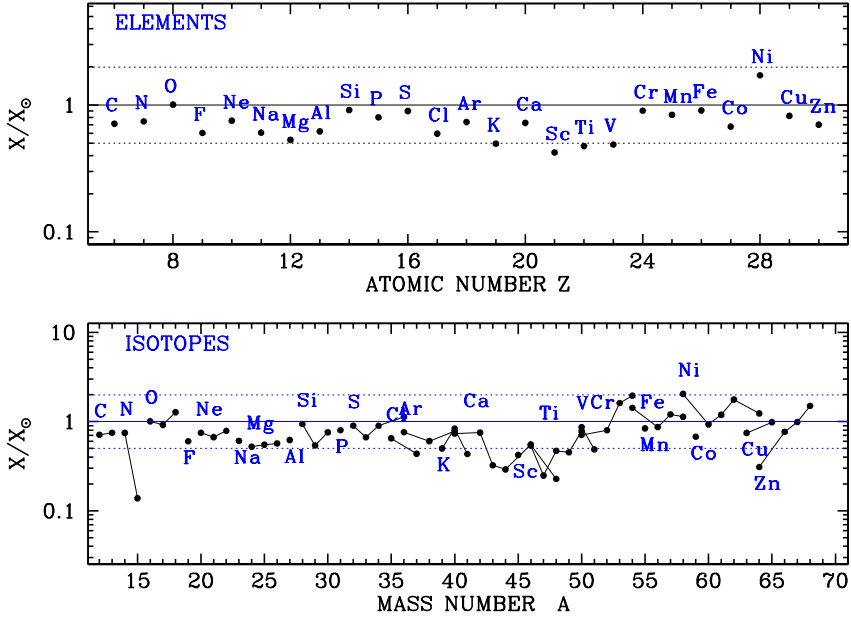}
\caption{Abundances of the chemical evolution model of the solar neighborhood 4.5 Gyr ago, compared to solar abundances, from C to Zn. Elemental abundances ({\it top}) are co-produced within a factor of 2. Isotopic abundances ({\it bottom}) are also well reproduced, except $^{15}$N (for which novae -not included in this calculation - appear to be the source) and some Ca and Sc isotopes. Note that mono-isotopic F is produced by neutrino-nucleosynthesis in massive stars, according to the calculations of Woosley and Weaver (1995) adopted here. More than half of solar Fe results from SNIa (with yields adopted form Iwamoto et al. 1999), while part of C and N comes from intermediate mass stars (with yields adopted from van den Hoek and Gronewegen 1997).}
\end{center}
\end{figure}

With those parameter adjustments, it remains to be seen whether the other observables of the system (Sec. 3.1) are reproduced. In Fig. 15 it is seen that the 
current model SFR is well within observational uncertainties. Note that the derived SFR history is rather flat, around an average value of $\sim$3.8 \ms/yr (alternatively, it could be fitted with a broad gaussian).

The resulting age-metallicity relationship (left bottom panel in Fig. 15) fits the data approximately, but its should be stressed that uncertainties in stellar ages are fairly large; in fact, dispersion in the age-metallicity relation is even larger than displayed in Fig. 15, to the point that this observable may be of very little use as a constraint for the local GCE (see Sec. 3.1.4 and Fig. 18).

The rise of Fe/O (right panel of Fig. 15) is due to delayed contribution of Fe by SNIa. The adopted SNIa rate is from Greggio and Renzini (1983) and it is assumed that 0.04 of the binaries produce SNI; the resulting SNIa rate as a function of time appears on the left panel. 

Finally, the local GCE model, combined with the adopted stellar yields, should also reproduce the pre-solar composition, well established after meteoritic and photospheric measurements (e.g. Lodders 2003). The results of such a comparison (model results at time $T$-4.5 Gyr divided by pre-solar composition) are displayed in Fig. 16. It can be seen that all elements and almost all isotopes are nicely co-produced (within a factor of 2 from their pre-solar value), with key elements such as O and Fe being very well reproduced. Taking into account the large abundance variations
between O and Sc (a factor of 10$^6$) this agreement should be considered as a triumph for stellar nucleosynthesis models. Some exceptions to this achievement are noted in the legend of Fig. 15, but the overall result is quite encouraging:
the adopted IMF, SFR  and stellar yields reproduce very well the solar system composition (assumed to be typical of the local ISM 4.5 Gyr ago).

\subsubsection{The local evolution of Deuterium}

Modelling the GCE of deuterium is a most straightforward enterprise, since this fragile isotope is 100\% destroyed in stars of all masses (already on the pre-main sequence)
 and has no known source of substantial production other than Big Bang nucleosynthesis (BBN). 
 In the framework of the simple Closed box model with IRA, its abundance is given by:
 \begin{equation}
 D \ = \ D_P \ \sigma^{{R}\over{1-R}}
 \end{equation}
 where $D_P$ is the primordial abundance,  $\sigma$ the gas fraction and $R$ the return fraction. 
 If the boundary conditions of its evolution (namely the primordial abundance $D_P$  and the present day one) were precisely known, the degree of {\it astration}, i.e the degree of processing of galactic gas inside stars, which depends on the adopted IMF and SFR, would be severely constrained. This, in turn, would be another  strong constraint (not appearing in the list of Sec. 3.1) for models of local GCE. For typical values of $\sigma\sim0.2$ and $R\sim$0.3 one obtains: $D/D_P\sim$0.5 in the Closed box model, i.e. depletion by a factor of $\sim$2. However, the true depletion factor should be smaller, because  infall (of presumably primordial composition) is required for the GCE of the local disk, as argued in the previous section.

\begin{figure*}
\centering
\includegraphics[width=0.55\textwidth,angle=-90]{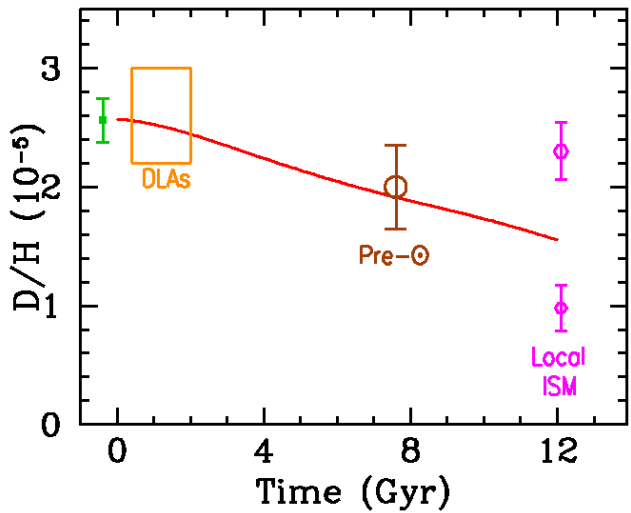}
\caption{Evolution of deuterium in the solar neighborhood, as a function of time. The adopted   model satisfies all major local observational constraints (see Fig. 15). Data (with error bars or inside box) correspond to: primordial $D_P$ (resulting from standard Big Bang nucleosynthesis + baryonic density inferred from analysis of the cosmic microwave background), observations of high redshift gas clouds (DLAs), pre-solar (inferred from observations of stellar wind composition) and in the local ISM (through different interpretations of UV data from the FUSE satellite). } 
\label{Fig3}
\end{figure*}

The primordial abundance of D is now well determined, since observations of D in high redshift gas clouds agree with abundances derived from observations of the Cosmic Microwave Background combined to  calculations of standard BBN (see e.g. Prantzos 2007 and reference therein). However, the present day abundance of D in the local ISM is now under debate. Indeed, UV measurements of the FUSE satellite  along various lines of sight suggest substantial differences (a factor of two to three) in D abundance between the Local Bubble and beyond it. Until the origin of that discrepancy is found (see Hebrard et al. 2005 and Linsky et al. 2006), the local GCE of D in the past few Gyr will remain poorly understood: naively, one may expect that a high value would imply strong late infall of primordial composition, while a low value would imply strong late astration. In any case, corresponding models should also satisfy all other local observables, like the overall metallicity evolution and the G-dwarf metallicity distribution, which is not an easy task.

\subsubsection{Uncertainties in the evolution of the Solar neighborhood}

\begin{figure}
\begin{center}
 \includegraphics[width=6cm,angle=-90]{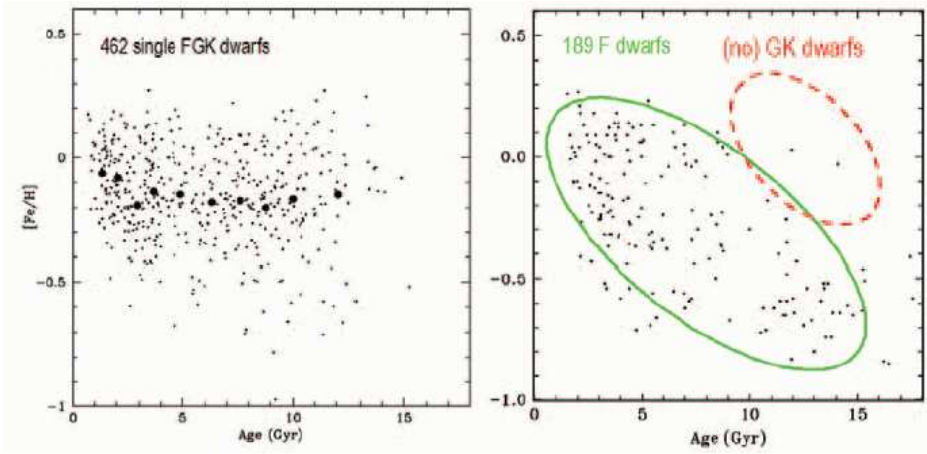}
\caption{Age-metallicity relationship in the solar neighborhood. {\it Left}: data from Nordstrom et al. (2004) for FGK stars within 40 pc. Thick dots indicate average metallicities in the corresponding age bins; no age-metallicity relation appears in those data. {\it Right:} data for 189 F dwarfs from the study of Edvardsson et al. (1993), suggesting an age-metallicity relation; by construction, however, old and metal-rich stars are excluded from that sample. Figure from Nordstrom et al. (2004).}
\end{center}
\end{figure}

Our understanding of the evolution of the solar neighborhood depends on how well are established the relevant  observational constraints (listed in Sec. 3.1).
It has recently been suggested that some of those constraints may be less well understood than thought before.
 
Stellar ages are much harder to evaluate than stellar metallicities, and the form of the local AMR has varied considerably over the years. The seminal work of Edvardsson et al. (1993) on 189 F-dwarfs established  a clear trend of decreasing metallicity with age, albeit with substantial scatter (Fig. 18 right). 
Such a trend is compatible with (and predicted by) all simple models of local GCE, either closed or open (i.e. with infall) models. It should be noted, however, that the adopted selection criteria in that paper introduced a bias against old metal-rich and young metal-poor  stars.

The large survey  of Nordstrom et al. (2004), concerning $\sim$14000 F and G stars with  3-D kinematic information (but less precise spectroscopy than the Edvardsson et al. study), provides a radically different picture: the volume limited subsample of 462 stars with "well-defined" ages withing 40 pc displays a flat AMR (an average metallicity of [Fe/H]$\sim$-0.2 at all ages) with a very large scatter (Fig. 18 left). Acounting for the fact that the oldest stars have the largest age uncertainties does not modify the flatness of the AMR.
If confirmed, a flat AMR would require different assumptions than those adopted in current models (e.g. substantial late infall to dilute metals, with a strong impact on D evolution or on the metallicity distribution).

\begin{figure}
\begin{center}
\includegraphics[angle=-90,width=0.46\textwidth]{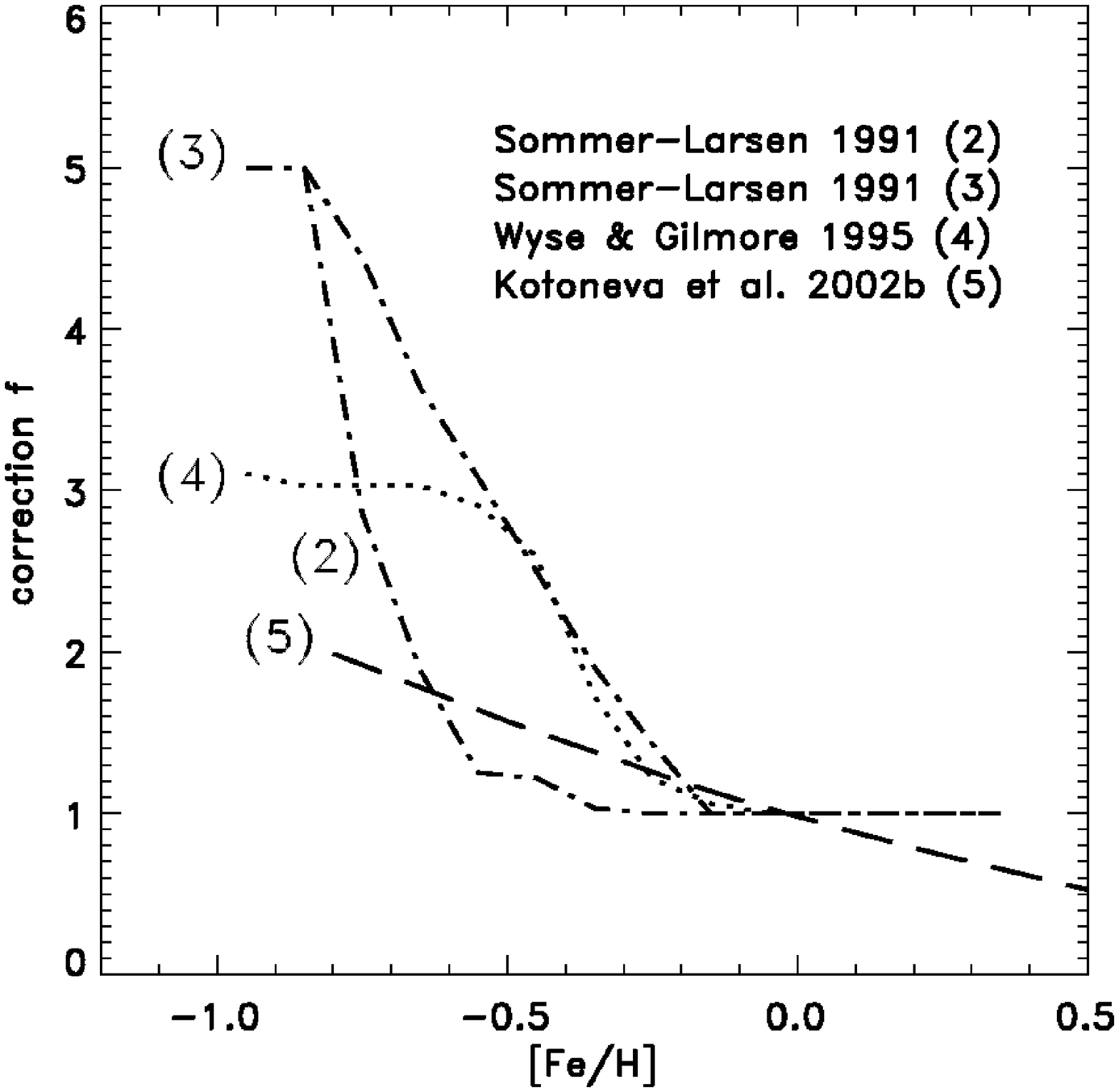}
\qquad
\includegraphics[angle=-90,width=0.46\textwidth]{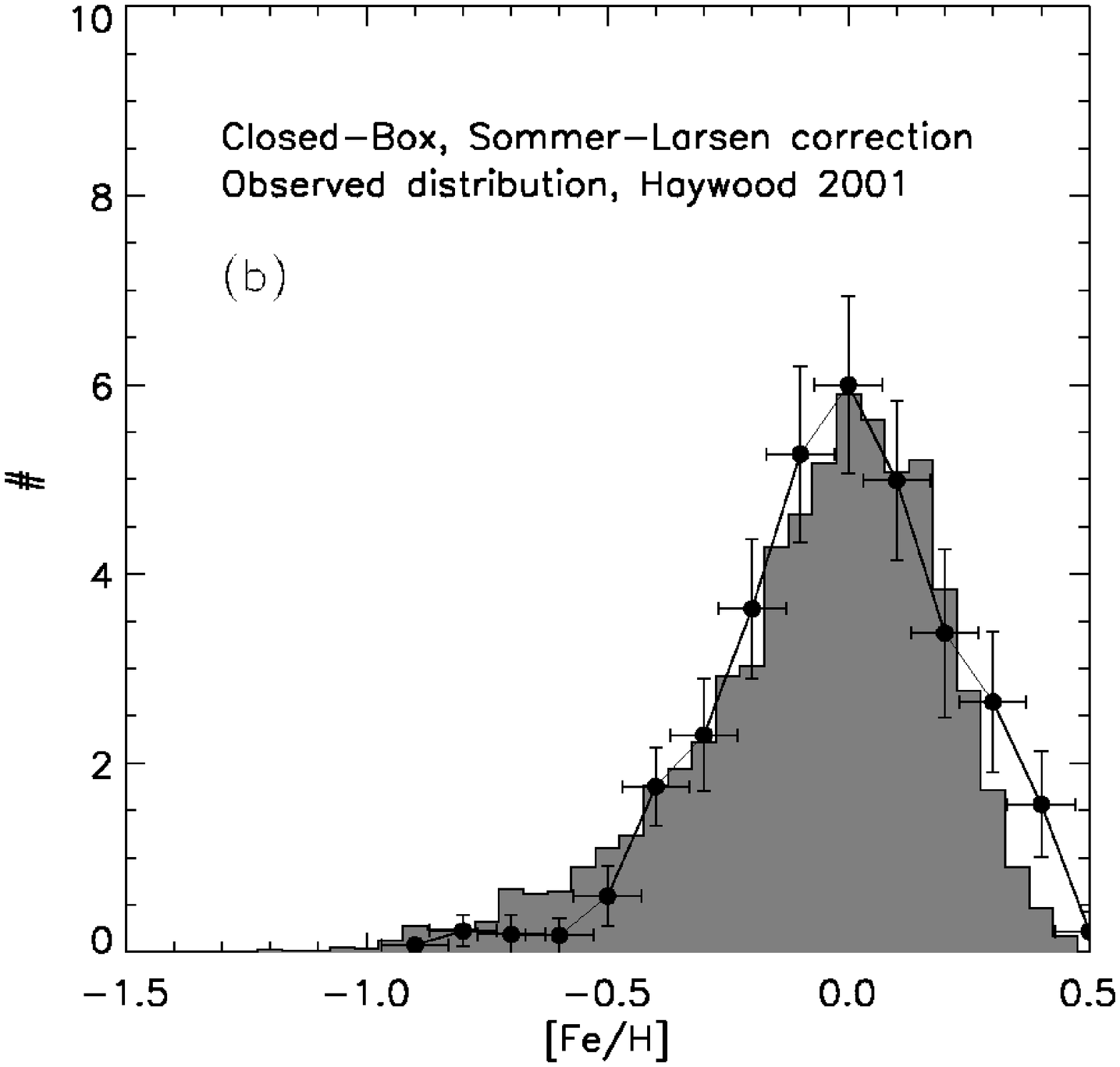}
\caption{{\it Left:} Correcting factors as a function of metallicity, to be applied to the local (thin disk) metallicity distribution, according to Haywood (2006); they are based on estimates of velocity dispersion as a function of metallicity. {\it Right:} If (the inverse of) those correction factors are applied to the results of a closed box model, the resulting metallicity distribution (grey shaded aerea) ressembles closely the observed one (from Haywood 2006).}
\end{center}
\end{figure}

The form of the second "pillar" of local GCE, namely the local MD, has also been revisited recently by Haywood (2006). He argues that, since the  more metal-poor stellar populations have larger dispersion velocities vertically to the disk,  larger scaleheight corrections should be applied to their numbers in order to get their true surface density. By adopting such steeply decreasing correction factors with metallicity (Fig.  19 left) he finds then that a closed box model nicely account for the corrected MD (Fig. 19 right). He notes that the (volume corrected) local population at -1$<$[Fe/H]$<$-0.5 comprises 15-20 \% of the total; it should then be ``naturally'' considered as the thick disk population, which weights $\sim$7 M$_{\odot}$ pc$^{-2}$, compared to  $\sim$35 M$_{\odot}$ pc$^{-2}$ for the total stellar population in the solar cylinder.

The relationship of the thick and thin disks is not yeat clear; if thick disk stars were mostly accreted from merging/disrupted satellites, that population cannot be considered as belonging to the early phase of the disk and the arguments of Haywood (2006) do not hold. Also, in view of the large scaleheight of the thick disk (1400 pc)  and of  the elliptical orbits of its stars (vs. circular orbits for those of the thin disk) one may wonder how large the local ``chemical box'' could be and still be considered as a single system with well defined evolution. It is clear, however, that corrections to the observed local MD should be carefully considered before comparing to GCE models and this would certainly impact on the required infall timescales (which could be smaller than the ``canonical'' value of $\sim$7 Gyr currently adopted).

\subsection{The Milky Way halo}

The list of observational constraints for the Galactic halo is much shorter than for the solar neighborhood. In particular, we have no information on the gas fraction at the end of its evolution, no isotopic abundanecs available  and there is no age-metallicity relationship. Age differences  of Galactic globular clusters  are evaluated to $\sim$1-2 Gyr (e.g. Salaris and Weiss 1993) and this is usually taken to be the duration of halo formation. Note that this timescale is much longer than the $\sim$10$^8$ yr derived in the monolithic collapse scenario of Eggen et al. (1962).

The two most important pieces of information on the Galactic halo are: (i) the abundance patterns of various elements and (ii) the metallicity distribution.

\subsubsection{Halo star composition: any signatures of  Pop. III stars ?}

 Detailed abundance patterns have now been established in the case of halo stars, for the majority of the chemical elements (see Hill, this volume).
 In the case of heavier than Fe elements, their interpretation is hampered at present by our poor understanding of the production sites of the s- and r- nuclides (see Arnould, this volume). In the case of intermediate mass elements, up to the Fe-peak, the situation is more favourable than in the case of the local disk, since the system is thought to be enriched only by the products of massive stars and not of SNIa (which operate effectively on longer timescales than $\sim$1 Gyr). Observed abundance patterns of those elements constrain then corresponding massive star yields (folded with a stellar IMF). In particular, the composition of the most metal-poor stars (say, below [Fe/H]$\sim$--3) constrains the yields of the first massive star generations (see also Limongi, this volume), those formed from material of primordial composition (without metals) and generically called {\it Pop. III stars}.

On theoretical grounds, it has been conjectured that Pop. III stars
should be more massive on average than stars of subsequent, metal
enriched, generations. The reason is that, in the absence of metals,
gas cannot radiate efficiently away its heat (the most efficient cooling agent
in that case being the H$_2$ molecule); thus, it cannot
reduce sufficiently its  internal pressure and collapse gravitationally.
Even in the absence of metals,
the complexity of the problem of gas collapse and fragmentation 
(involving various forms of instabilities, turbulence, magnetic fields, 
ambipolar diffusion etc.) is such that the typical mass of
the first stars is very poorly known at present
(see Glover 2005 and references therein). It may be as ``low'' as the
mass of a typical massive O star today (i.e. a few tens of \ms)
or as large as several hundreds of \ms. It is even possible that, depending
on ambient density, two ranges of typical masses co-exist for Pop. III stars,
i.e. one around a few \ms, and another around a few hundreds of \ms \ (Nakamura
and Umemura 2002). Is there any evidence from current observations for or against those theoretical ideas ?

\begin{figure}
\begin{center}
 \includegraphics[width=10cm,angle=-90]{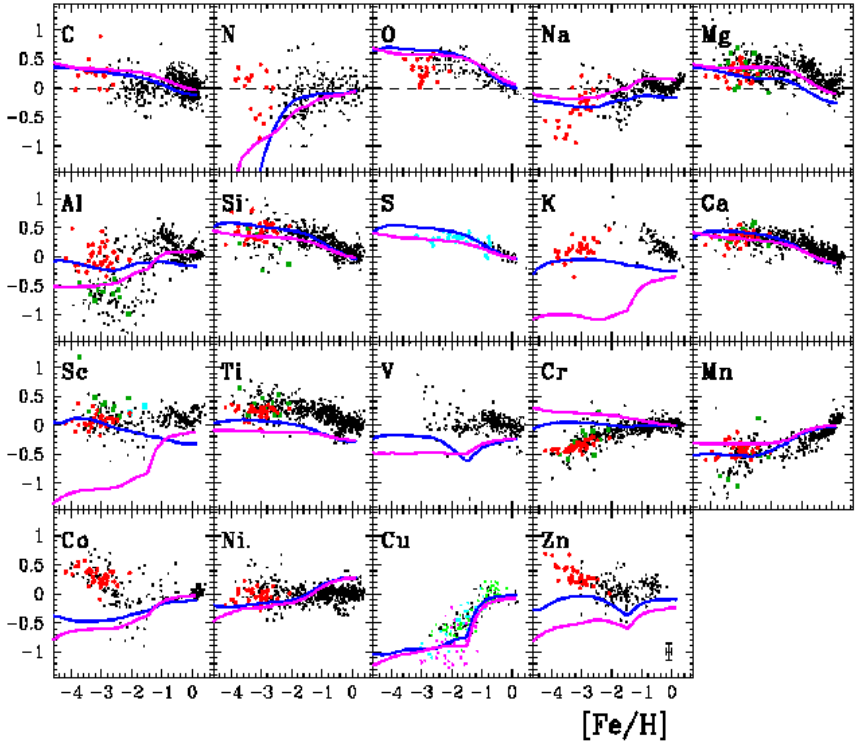}
\caption{Abundance ratios [X/Fe] as a function
of metallicity [Fe/H] in stars of the Milky Way; small data points
are from various sources, while the large data points at low 
metallicity are from the  VLT survey of Cayrel et al. (2004).
Results of standard chemical evolution models, performed with two
sets of metallicity dependent massive star yields 
(Woosley and Weaver 1995, {\it solid curves} ; 
Chieffi and Limongi 2004, {\it dashed
curves}) are also displayed. Yields for SNIa are adopted from 
Iwamoto et al. (1999) and for intermediate mass stars from 
van den Hoek and Gronewegen  (1997). While the behaviour of alpha elements
and Mn is correctly reproduced (at least qualitatively), 
there are large discrepancies
in the cases of Cr, Co and Zn.}
\end{center}
\end{figure}

A summary of observational data for intermediate mass elements in halo stars appears in Fig. 20, where comparison is made with a GCE model (from Goswami and Prantzos 2000). The adopted stellar yields (from Woosley and Weaver 1995 and Chieffi and Limongi 2004) are from non-rotating Core Collapse supernovae (CCSN) in the ``normal'' mass range
of 12 to $\sim$40 \ms, exploding with ``canonical'' energies of $E=$10$^{51}$erg and with progenitor metallicities from Z=0 to
Z=\zs. The salient features of that comparison can be summarized as follows:
\begin{itemize}
\item{The $\alpha$/Fe ratio (where $\alpha$ stands for alpha elements like
O, Mg, Si, Ca) remains $\sim$constant$\sim$2-3 times solar from [Fe/H]$\sim$--1 down to the lowest observed metallicities. This is attributed to SNII, while its solar value is attributed to the later contribution of Fe by SNIa.}
\item{In the adopted stellar models (no mass loss or rotation) N is produced as secondary, while observations show that it behaves as a primary (i.e. N/F is constant); this point will be discussed in the next section.}
\item{Abundances of elements affected by the {\it odd-even effect} (see Sec. 2.3.2), like Na and Al, require uncertain and potentially large NLTE corrections and it is difficult, at present, to use them in order to constrain stellar yields.}
\item{There is a group of elements below the Fe-peak, including K, Sc, Ti and V, the abundance pattern of which is poorly described by the adopted yields (as well as by those of the Tokyo group of Nomoto et al. 2006);  this apparently "generic" problem of current stellar models is discussed by Limongi (this volume).}
\item{The decreasing trend of Mn/Fe and Cu/Fe with decreasing metallicity is in
nice qualititative agreement with theoretical expectations.}
\item{The Fe-group nuclei Cr, Co and Zn behave very differently than expected on theoretical grounds.}
\end{itemize}

The reasons for the latter discrepancy probably lie in our poor understanding of
the explosion mechanism and of the nature of the early CCSN.
A large amount of work was devoted in recent years to understanding the peculiar abundance patterns among Fe-peak elements observed in extremely metal-poor (EMP) stars (for reviews see  Cayrel 2006 or Limongi, this volume, and references therein). In those works it was explored whether some specific 
property of the Pop. III stars (characteristic mass, rotation, explosion energy 
etc.) was quite different from the corresponding one of their more metal
rich counterparts and produced a different nucleosynthetic
pattern in the ejecta.
Two main classes of solutions (not mutually exclusive) appear to arise from those works:

\begin{figure}
\begin{center}
\includegraphics[angle=-90,width=0.46\textwidth]{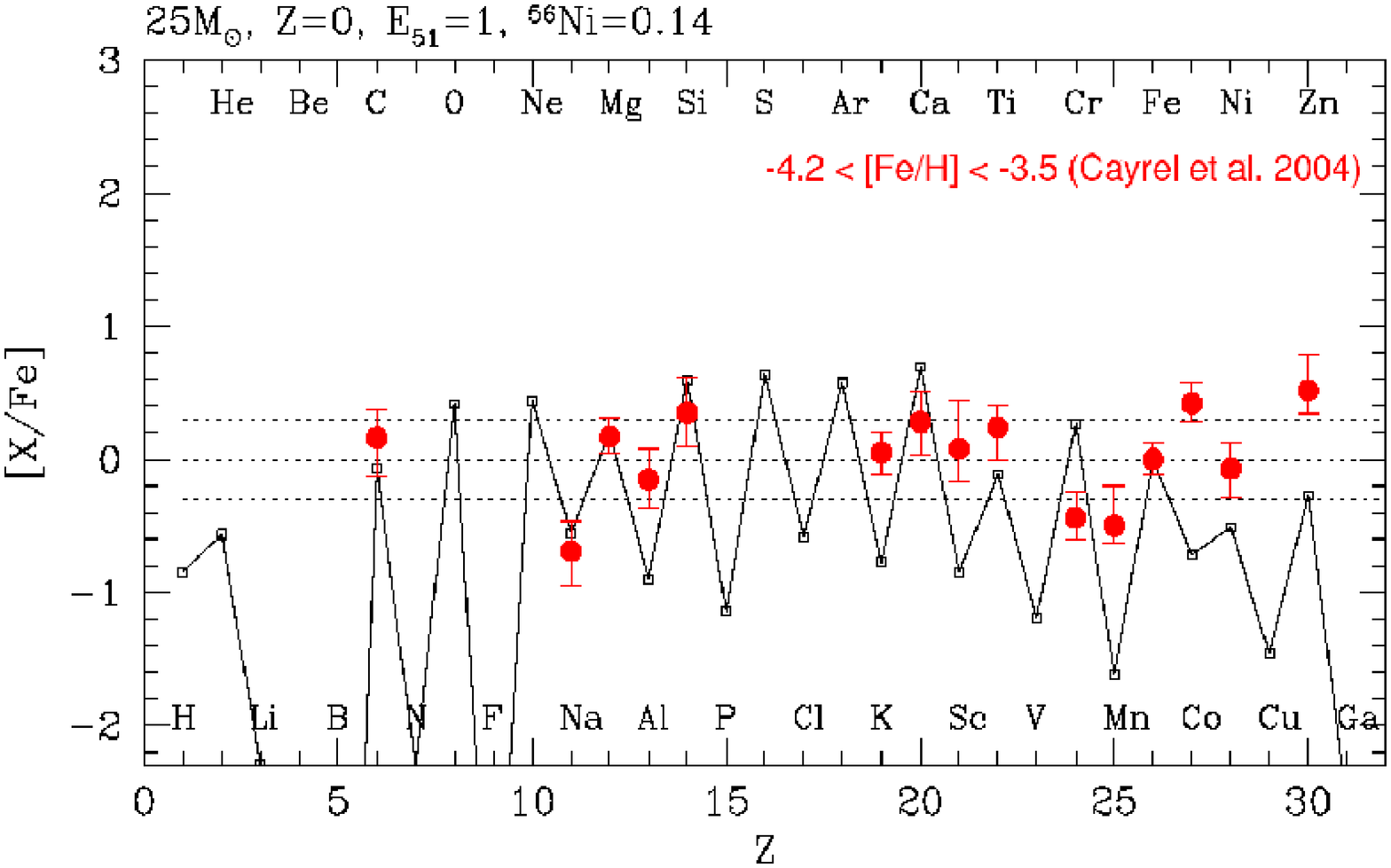}
\qquad
\includegraphics[angle=-90,width=0.46\textwidth]{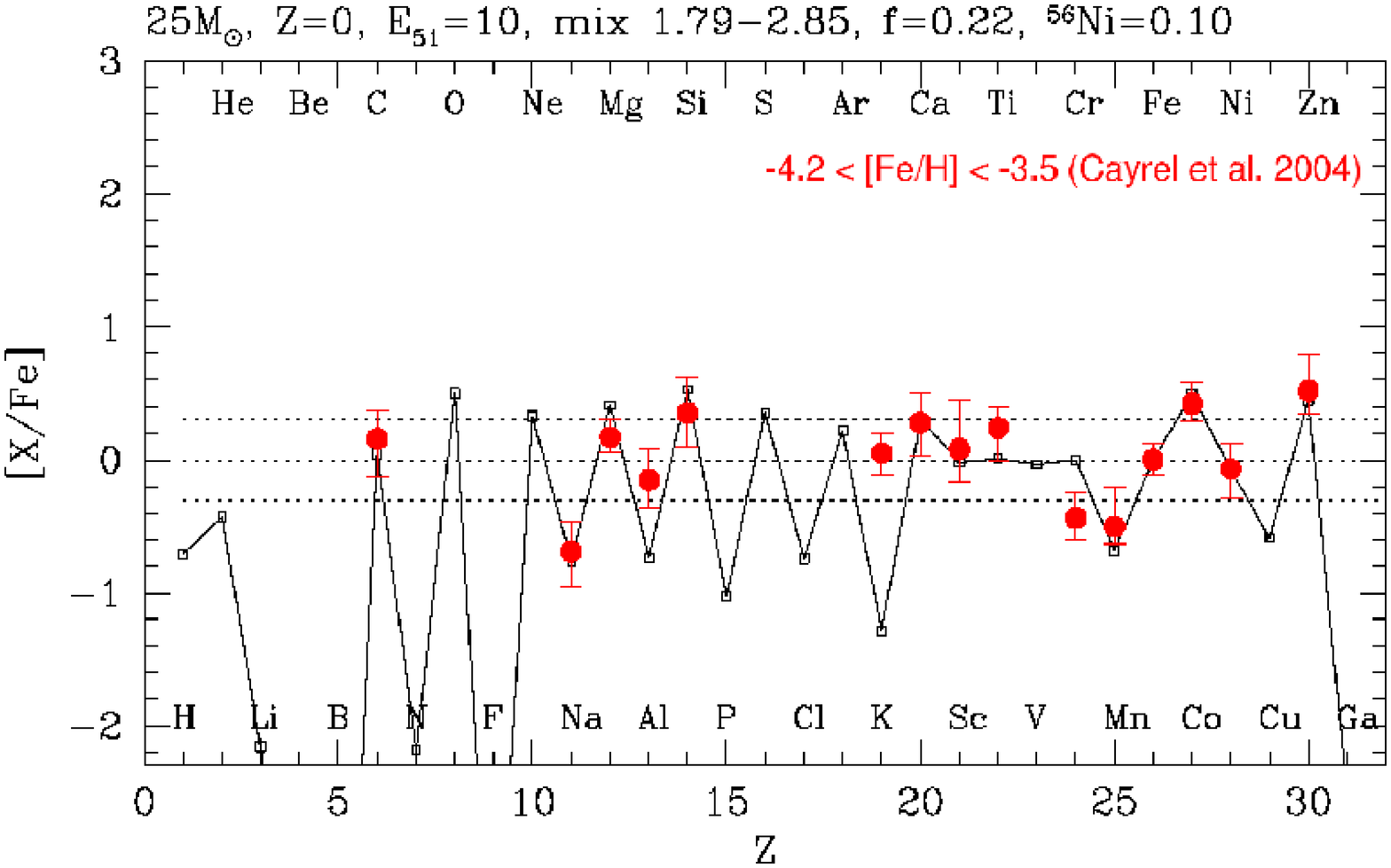}
\caption{Nucleosynthesis yields of a 25 \ms \ star, compared to observations of extremely metal-poor stars (vertical error bars, from Cayrel et al. 2004). {\it Left:} The star explodes with a canonical energy of 1.5 10$^{51}$ erg; {\it Right}: The star explodes as a hypernova, with an energy of 10 10$^{51}$ erg (both figures from Nomoto et al. 2006).   Abundance ratios in the Fe-peak are much better reproduced in the second case.}
\end{center}
\end{figure}

1) {\it Normal massive stars (10-50 \ms)}, exploding as core collapse SN (CCSN) and leaving behind a neutron star or a black hole. It was shown (essentially by the Tokyo group, see Nomoto et al. 2006 for a recent review) that {\it higher energies} than the canonical one of $E=10^{51}$ ergs  {\it combined with asphericity}
of the explosion can help to improve the situation concerning the Zn/Fe and Co/Fe ratios observed in EMPs (Fig. 21 right) . Material ejected along the rotation  axis (in the form of jets) has high entropy
and is found to be enriched in products of $\alpha$-rich freeze-out (Zn and Co), as well as Sc (which is 
generically undeproduced in spherical models); this kind of models seems at present
the most promising, but this is not quite unexpected (since they have 
at least one more degree of freedom w.r.t. spherical models). Such properties (high energies, asphericity) are indeed observed in the local Universe for {\it hypernovae}, a class of energetic CCSN. However, hypernovae are not so frequent today as to affect the later chemical evolution of galaxies (normal CCSN plus SNIa can easily account for e.g. the solar abundance pattern). If hypernovae indeed affected the early chemical evolution, they must have been much more abundant in  Pop. III than today. No satisfactory explanation for such a large hypernova fraction in the early Universe exists up to now. It seems hard to conceive that the typical energy of the explosion was substantially higher in Pop. III stars (all other parameters, except metallicity, being kept the same). On the other hand, rotation effects may indeed be stronger in low metallicity environments, because of smaller angular momentum losses due to lower mass loss rates (e.g. Meynet, this volume). In any case, none of the models explored so far appears to account for the early trend of Cr/Fe, which remains a  mystery at present.

\begin{figure}
\begin{center}
\includegraphics[angle=-90,width=0.43\textwidth]{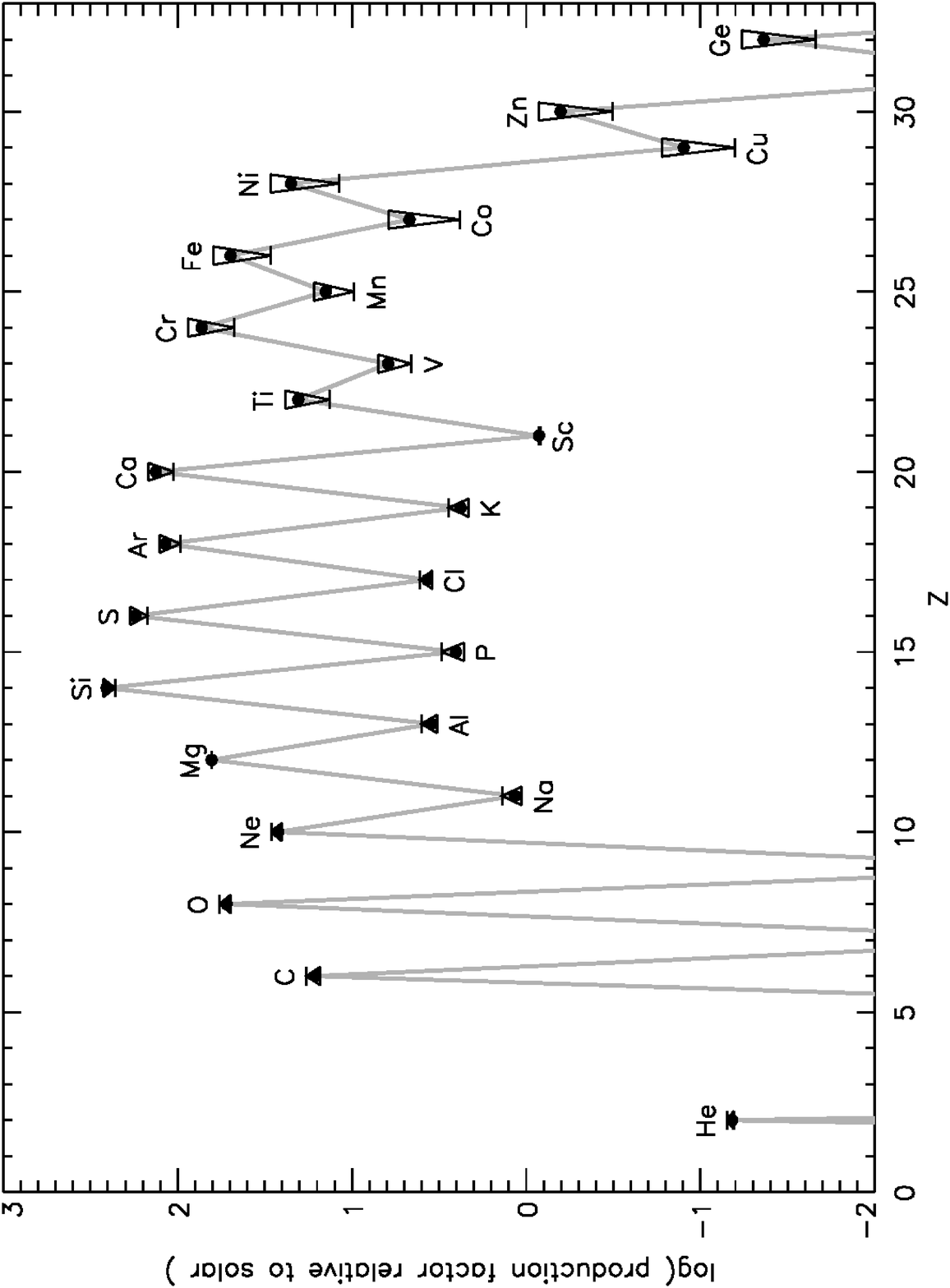}
\qquad
\includegraphics[angle=-90,width=0.5\textwidth]{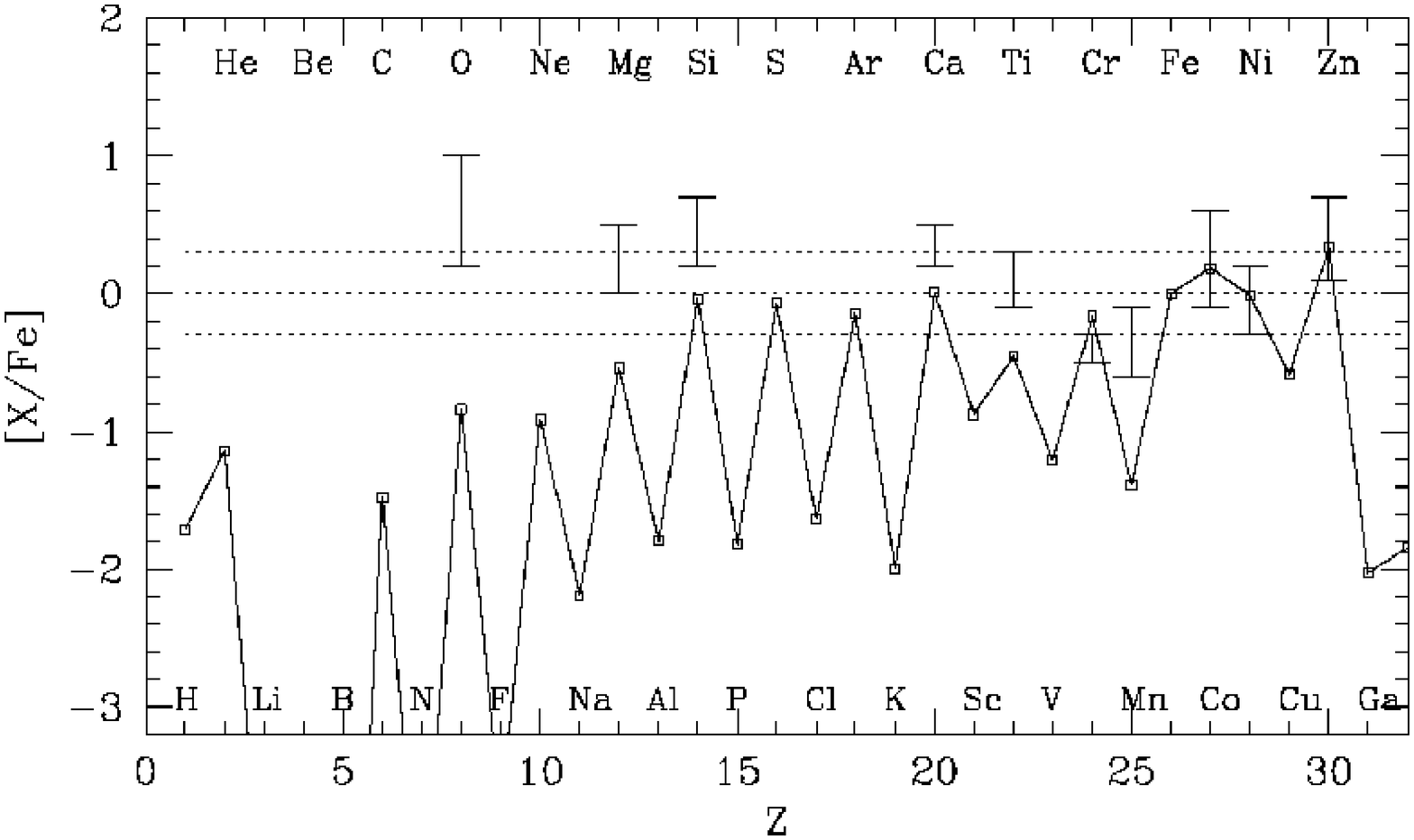}
\caption{Nucleosynthesis yields of Very Massive Stars (VMS, with mass $>$100 \ms). {\it Left:}  Stars in the 140 to 260 \ms \ range  explode as Pair Instability Supernovae (PISN) and do not produce enough Zn (from Heger and Woosley (2002). {\it Right}: Stars in the 300-1000 \ms \ range explode after core collapse and produce significant amounts of Zn (from Ohkubo  et al. 2006).   Abundance ratios in the Fe-peak are much better reproduced in the second case.
A Pop. III population composed exclusively of VMS (in the 140-1000 \ms \ range) and with an appropriate IMF is obviously compatible with observations of EMP stars.}
\end{center}
\end{figure}

2) {\it Very massive stars (VMS)}, above 100 \ms. Such stars are thought to: collapse to black holes, if M$<$140 \ms; explode during oxygen burning as pair-implosion SN (PISN), if 140$<$M/\ms$<$300; and again collapse without ejecting metals if M$>$300 \ms \ (see Heger and Woosley 2002 and references therein). In the case of PISN, it was shown that they do not produce enough Zn to account for the oberved high Zn/Fe ratio in EMP stars (Heger and Woosley 2002, Umeda and Nomoto 2006 and Fig. 22 left) and they were thus excluded as major contributors to the early chemical evolution of the MW. This shortcoming cast doubt as to the existence of a Pop. III generation composed exclusively of VMS. However, in a recent work Ohkubo et al. (2006) explore the conditions under which {\it rotating stars} of M$=$500  and 1000 \ms \ {\it do explode and eject metals} (although a massive black hole is ultimately produced in both cases). They find explosions for a certain region of their parameter space and substantial production of Zn as a generic feature (see Fig. 22 right). Even if each class of PISN or 500-1000 \ms \ stars cannot reproduce the observed EMP abundance pattern alone, it is obvious that considering the full 140-1000 \ms \ range (folded with an appropriate IMF) will account for such  a pattern. Thus, {\it VMS cannot be excluded at present as candidates for Pop. III stars}, at least not on ``chemical'' grounds.

\begin{figure}
\begin{center}
 \includegraphics[width=7cm]{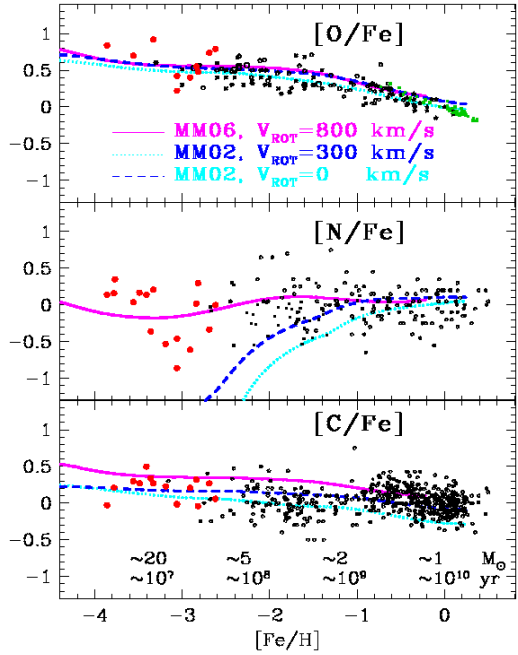}
\caption{Evolution of oxygen, nitrogen and carbon in the Milky Way (from top to bottom). Observations are from various sources, while model results correspond to three sets of the Geneva stellar yields: stars with no rotation, with rotational velocity $V$=300 km/s (as observed in the solar neighborhood) at all metallicities, and for $V$=800 km/s for low metallicity stars. The latter case provides the required primary N to fit the low metallicity data (see text). Timescales corresponding to metallicity and masses of stars dying on such timescales appear in the bottom panel.}
\end{center}
\end{figure}

\subsubsection{The quest for primary Nitrogen}

The behaviour of N as primary (i.e. [N/Fe]$\sim$0) was known for sometime, but it was recently confirmed from VLT mesurements (Spite et al. 2005) down to the realm of very low metallicities ([Fe/H]$\sim$-3, see Fig. 23 middle). For a long time, the only known source of primary N was Hot Bottom Burning (HBB) in AGB stars (see Sec. 2.3.4). Even the most massive AGBs ($\sim$8 \ms) have lifetimes ($\sim$5 10$^7$ yr) considerably longer than those of typical SNII progenitors (20 \ms \ stars living for $\sim$10$^7$ yr) and it is improbable that they contributed to the earliest enrichment of the Galaxy\footnote{Note, however, that the timescales of the early Galactic evolution are not constrained (there is no age-metalicity relation) and the contribution of AGBs even as early as [Fe/H]$\sim$--3 cannot be absolutely excluded.}. On the other hand, massive stars were though to produce N only as secondary (from the {\it initial} CNO) and not to be at the origin of the observed behviour.

Rotationally induced mixing in massive stars changed the situation considerably: N is now produced by H-burning of C and O {\it produced inside the star}\footnote{N is produced after mixing of protons in He-rich zones, where $^{12}$C is produced from the 3-$\alpha$ reaction.}, i.e. as primary.  The first set of models, rotating at 300 km/s at all metallicities, did not provide enough primary N at low metallicities to explain the data (see Fig. 23). Assuming that low metallicity massive stars were rotating faster than their high-metallicity present-day counterparts  (at 800 km/s) leads to a large production of primary N, even at low $Z$ (see Meynet, this volume) and allows one to explain the data (Fig. 23). Thus, there appears to be a "natural" solution to the problem of early primary N, which may impact on other isotopes as well (e.g. $^{13}$C, produced in a similar way).

\subsubsection{The early MW and  hierarchical galaxy formation}

In the previous sections, the early chemical evolution  of the MW was discussed independently of the cosmological framework in which it took place.
According to the currently dominant paradigm of hierarchical structure formation, the early phases of a galaxy's evolution are the most complex ones, as they involve multiple mergers of smaller sub-units. In the case of the Milky Way,  interesting "chemical signatures" of that period 
should still be left around us today, in the form of the metallicity distribution (MD) of long-lived stars .

The MD of Galactic halo field stars (HMD) is rather well known in the metallicity range --2.2$<$[Fe/H]$<$--0.8, while at lower metallicities its precise form has still to be established by ongoing surveys (Fig. 24 left). Its overall shape is well fitted  by a simple model of GCE as $dn/dlogZ \propto Z/p \ e^{-Z/p}$, where $p$ is the $yield$ of a stellar generation (Eq. 3.4); this function has a maximum for $Z=p$. The HMD peaks at a metallicity [Fe/H]=--1.6 (or [O/H]=--1.1, assuming [O/Fe]$\sim$0.5 for halo stars), pointing to a low yield $p$=1/13 of the corresponding value for the solar neighborhood. Such a {\it reduced} halo yield (see Eq. 2.34) is "classicaly" (i.e. in the monolithic collapse scenario) interpreted as  due to $outflow$ during halo formation (Hartwick 1976), at the large rate of 8 times the SFR (Prantzos 2003). How can it be understood in the framework of hierarchical merging ?

\begin{figure}
\begin{center}
\includegraphics[width=0.46\textwidth]{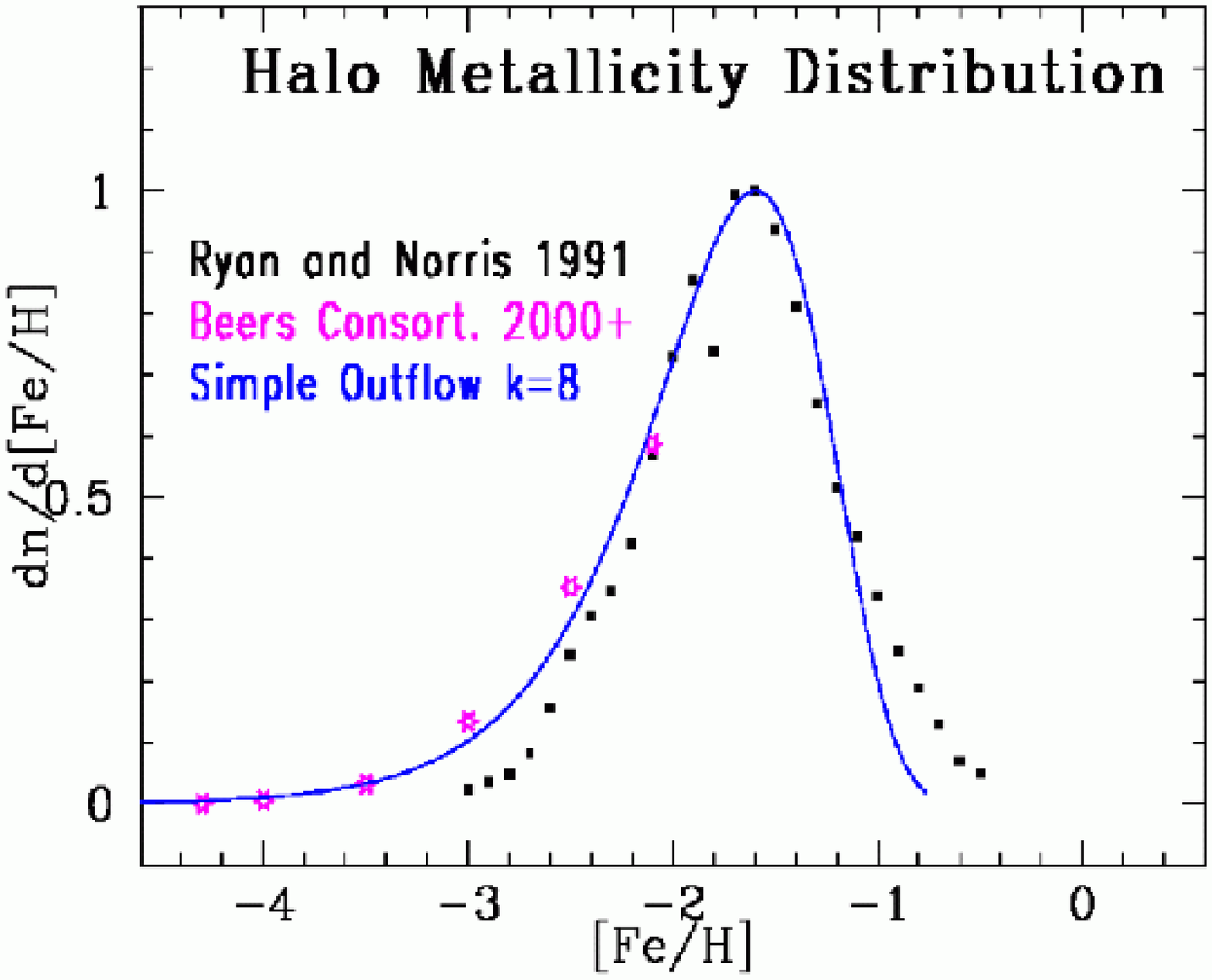}
\qquad
\includegraphics[width=0.46\textwidth]{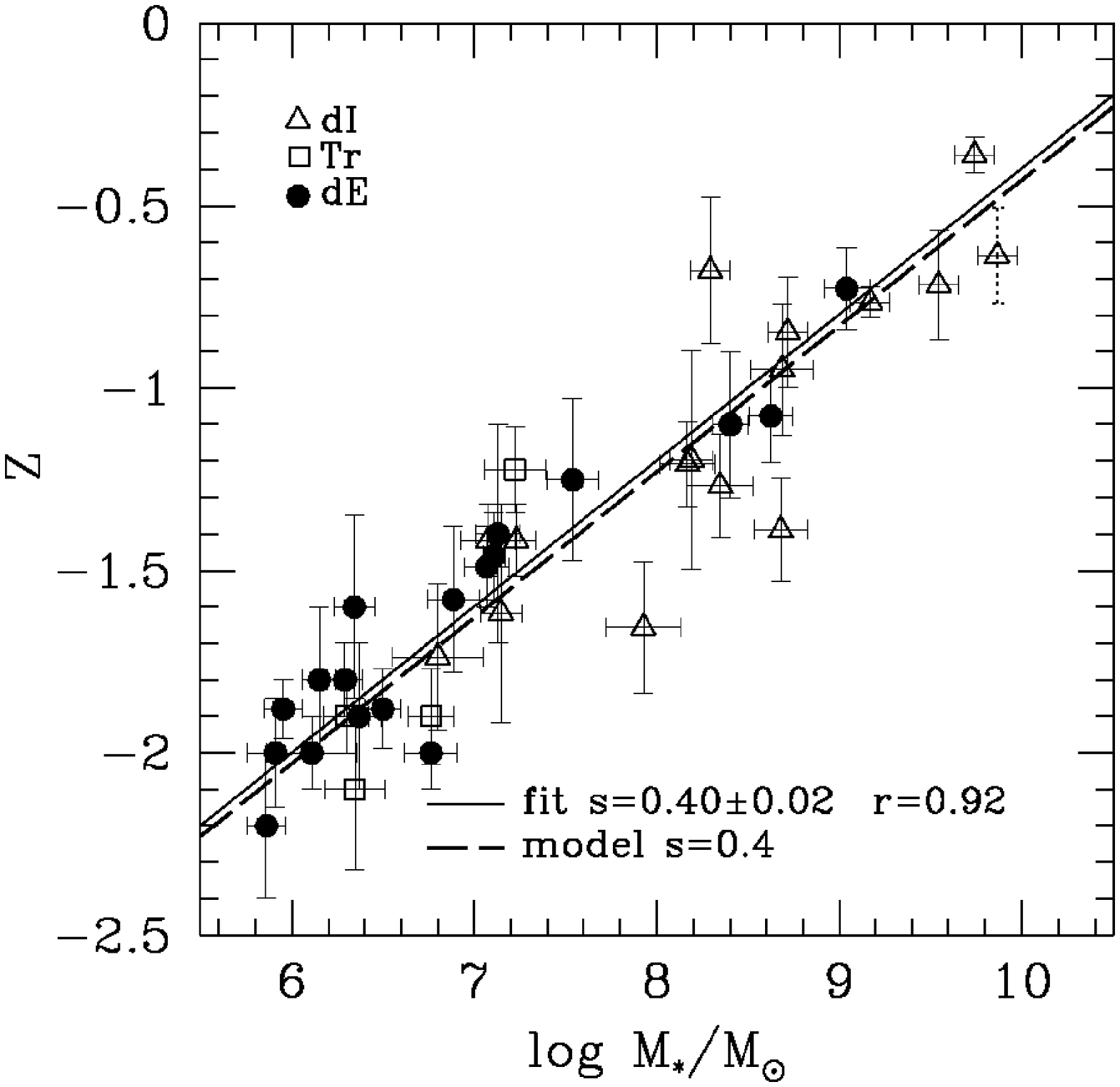}
\caption{{\it Left:} Metallicity distribution of field halo stars from Ryan and Norris (1991, dots), and the ongoing research of Beers and collaborators (T. Beers, private communication, asterisks). The curve is a simple model with outflow rate equal to 8 times the star formation rate. {\it Right:} Stellar metallicity vs stellar mass for nearby galaxies; data and model are from Dekel and Woo (2003). The MW halo, with average metallicity[Fe/H]=--1.6 (see left panel) or [O/H]=--1.1 and estimated mass 2 10$^9$ \ms \ falls below that relationship. }
\end{center}
\end{figure}

It should be noted that the typical halo metallicity ([Fe/H]=-1.6) is substantially lower, by more than 0.5 dex,  than the corresponding metallicities of nearby  galaxies of similar mass ($M_{Halo}\sim$2 10$^9$ \ms), as can be seen in Fig. 24 (right). That figure also displays the well known galaxian relationship between stellar mass and stellar  metallicity; most probably, that relationship results from mass loss, which is more important in lower mass galaxies, since the hot supernova ejecta escape more easily their swallow potential well (e.g.  Dekel and Woo 2003).

\begin{figure}
\begin{center}
\includegraphics[width=0.46\textwidth]{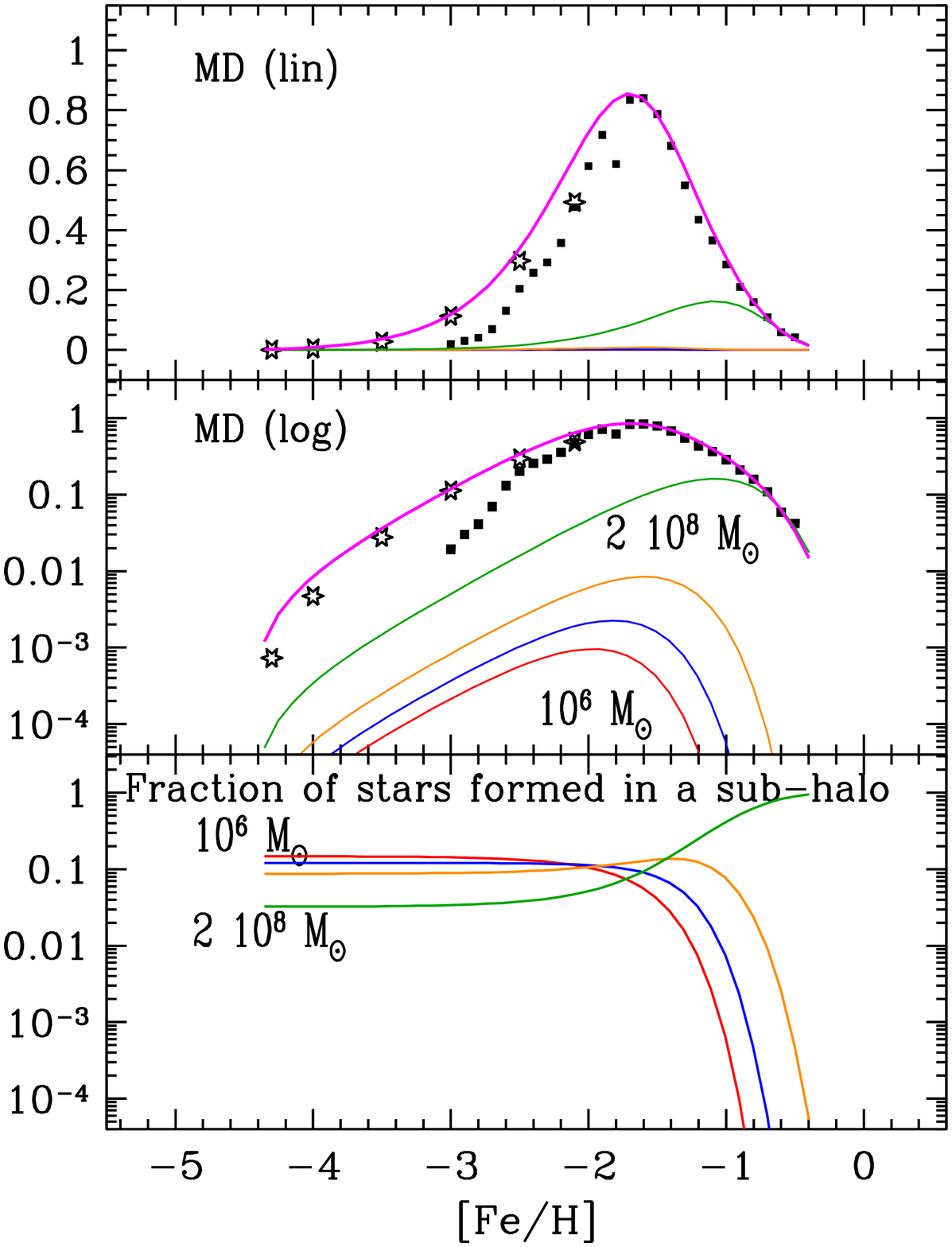}
\qquad
\includegraphics[width=0.46\textwidth]{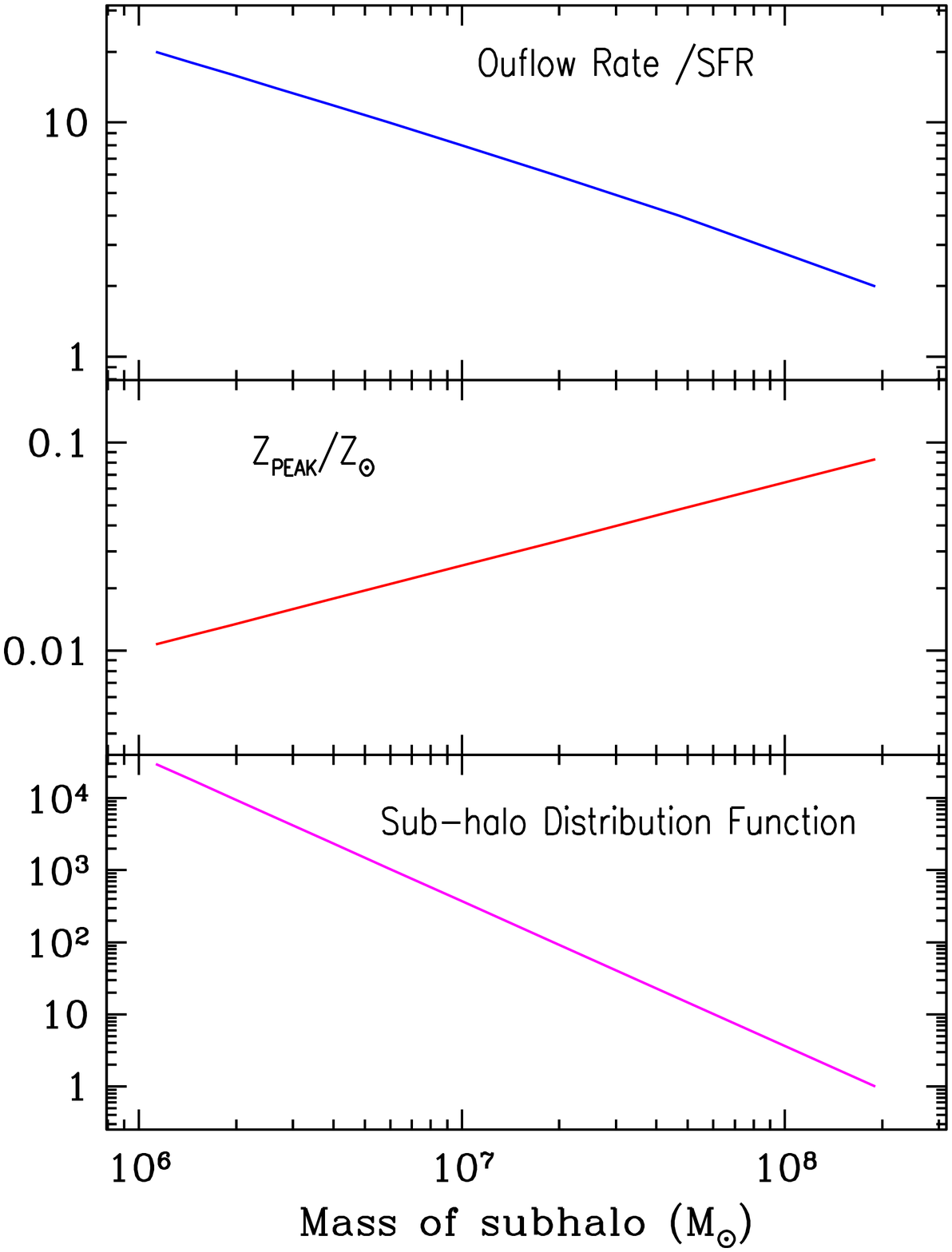}
\caption{{\it Left, top} and {\it middle} panels: Metallicity distribution (in lin and log scales, respectively) of the MW halo, assumed to be composed of a population of smaller units (sub-haloes). The individual MDs of a few sub-haloes, from 10$^6$ \ms \ to 2 10$^8$ \ms, are indicated in both panels (but clearly seen only in the middle), as well as the sum over all haloes (upper curves in both panels, compared to observations). Small sub-haloes contribute the largest fraction of the lowest metalicity stars ({\it bottom left}).  {\it Right panels:} Properties of the sub-haloes as a function of their mass.}
\end{center}
\end{figure}

Assuming that the MW halo has been assembled from sub-units similar to the low mass galaxies of Fig. 24 (right), one may interpret the HMD as the sum of  the MD of such low mass galaxies; it is assumed that each one of those galaxies (of mass $m$) evolved with an appropriate outflow rate and corresponding effective yield $p(m)=(1-R)/(1+k-R)$, where $k(m)$ is the outflow rate in units of the SFR and $R$ is the return mass fraction,  depending on the adopted stellar IMF. In that case, one has: HMD($Z$) = 1/$m_{Halo}$ $\int Z/p(m) \ e^{-Z/p(m)} \phi(m) m dm$, where $\phi(m)$ is the mass function of the sub-units  and $p(m)$  is the effective yield of each sub-unit (obtained directly from Fig. 24 (right)  as $p(m)=Z(m)$, i.e. smaller galaxies suffered heavier mass loss).

The results of such a simple toy-model for the HMD appear in Fig. 25 (top and middle left, from Prantzos 2006). The HMD is extremely well reproduced, down to the lowest metallicities, {\it assuming} $\phi(m) \propto m^{-2}$; such a halo mass function  results from recent high resolution numerical simulations for Milky Way sized dark haloes (Diemand et al. 2006, Salvadori et al. 2006). Low mass satellites (down to 10$^6$ \ms)  contribute most of the low metallicity stars of the MW halo, whereas the high metallicity stars originate in a couple of  massive satellites with $M\sim$ 10$^8$ \ms \ (Fig. 25 left, bottom).

Some properties of the sub-haloes  as a function of their mass appear also in Fig. 25 (right panels).  The outflow rate, in units of the corresponding SFR, is $k(m) = (1-R) (p_{True}/p(m) \ - \ 1)$, where $p_{True}$ is the yield in the solar neighborhood ($p_{True}=0.7 Z_{\odot}$, from the local G-dwarf MD).   
If the MW halo were formed in a a potential well as deep as those of comparable mass galaxies, then  the  large outflow rate required to justify the HMD ($k$=8) is puzzling; on the contrary, the HMD is readily understood if the MW halo is formed from a large number of smaller satellites, each one of them having suffered heavy mass loss according to the simple outflow model.

In summary, the halo MD can be understood in terms of hierarchical galaxy formation, {\it provided that each merging sub-halo behaves as described by the simple GCE  model with outflow}, the outflow rate depending on the sub-halo mass as suggested by observations of local dwarf galaxies.

\section{Conclusion}

In this tutorial, we present the formalism and the ingredients of standard GCE models. We also illustrate the concepts of GCE by applying simple models to the study of the local disk and of the Milky Way's halo, two systems with a large body of  available observational data. In both cases, useful information on those galaxian systems may be extracted from the data, interpreted in the framework of such simple models. It is stressed, however, that the history of the systems cannot be reconstructed in a  unique way, because (i) dispersion in the data does not allow it, and/or (ii) some of the data (e.g. the metallicity distributions) are independent of the form of the star formation rate.

It is clear that the future of the discipline lies in forthcoming chemodynamical
models of GCE, incorporating several physical ingredients and evolving in a cosmological framework. However, untill such models are properly validated/calibrated, simple GCE models such as those presented here will continue to be useful, allowing for a first, rapid interpretation of a large amount of observations.

\def\apj{{\it ApJ \ }}
\def\apjs{{\it ApJS \ }}
\def\aa{{\it A\&A \ }}
\def\aas{{\it A\&AS \ }}
\def\mn{{\it MNRAS \ }}
\def\ara{{\it ARAA \ }}

\end{document}